\documentclass[superscriptaddress,secnumarabic,nobibnotes,aps,
prd,showpacs,nofootinbib]{revtex4}
\usepackage{eurosym}
\usepackage{latexsym}
\usepackage{epsfig}
\usepackage{amssymb}

\newcommand{\ba}{\begin{eqnarray}}
\newcommand{\ea}{\end{eqnarray}}
\newcommand{\be}{\begin{equation}}
\newcommand{\ee}{\end{equation}}

\newcommand{\R}{\mathcal{R}}

\begin{document}

\title{Hybrid metric-Palatini stars}
\author{Bogdan Danil$\check{\mathrm{a}}$}
\email{bogdan.danila22@gmail.com}
\affiliation{Astronomical Observatory, 19 Ciresilor Street, Cluj-Napoca, Romania}
\author{Tiberiu Harko}
\email{t.harko@ucl.ac.uk}
\affiliation{Department of Physics, Babes-Bolyai University, Kogalniceanu Street,
Cluj-Napoca 400084, Romania,}
\affiliation{Department of Mathematics, University College London, Gower Street, London
WC1E 6BT, United Kingdom}
\author{Francisco S. N. Lobo}
\email{fslobo@fc.ul.pt}
\affiliation{Instituto de Astrof\'{\i}sica e Ci\^{e}ncias do Espa\c{c}o, Faculdade de
Ci\^encias da Universidade de Lisboa, Edif\'{\i}cio C8, Campo Grande,
P-1749-016 Lisbon, Portugal}
\author{M. K. Mak}
\email{mankwongmak@gmail.com}
\affiliation{Departamento de F\'{\i}sica, Facultad de Ciencias Naturales, Universidad de
Atacama, Copayapu 485, Copiap\'o, Chile}
\date{\today }

\begin{abstract}
We consider the internal structure and the physical properties of specific
classes of neutron, quark and Bose-Einstein Condensate stars in the recently
proposed hybrid metric-Palatini gravity theory, which is a
combination of the metric and Palatini $f(R)$ formalisms. It turns out that the theory
is very successful in accounting for the observed phenomenology, since it unifies
local constraints at the Solar System level and the late-time cosmic acceleration,
even if the scalar field is very light. In this paper, we derive the equilibrium
equations for a spherically symmetric configuration (mass continuity and
Tolman-Oppenheimer-Volkoff) in the framework of the scalar-tensor representation of the
hybrid metric-Palatini theory, and we investigate their solutions numerically for
different equations of state of neutron and quark matter, by adopting for the scalar
field potential a Higgs-type form. It turns out that the scalar-tensor definition of the potential can be represented as an Clairaut differential equation, and provides an explicit form for $f(\R)$ given by $f(\R) \sim \R + \Lambda_{\rm eff} $, where $\Lambda_{\rm eff} $ is an effective cosmological constant.
Furthermore, stellar models,
described by the stiff fluid, radiation-like, the bag model and the
Bose-Einstein Condensate equations of state are explicitly constructed in
both General Relativity and hybrid metric-Palatini gravity, thus allowing an
in-depth comparison between the predictions of these two gravitational
theories. As a general result it turns out that for all the considered
equations of state, hybrid gravity stars are more massive
than their general relativistic counterparts. Furthermore, two classes of
stellar models corresponding to two particular choices of the functional form of the scalar field (constant value, and logarithmic form, respectively), are also
investigated. Interestingly enough, in the case of a constant scalar field
the equation of state of the matter takes the form of the bag model equation
of state describing quark matter. As a possible astrophysical application of
the obtained results we suggest that stellar mass black holes, with masses
in the range of $3.8M_{\odot}$ and $6M_{\odot}$, respectively, could be in
fact hybrid metric-Palatini gravity neutron or quark stars.
\end{abstract}

\maketitle

\section{Introduction}

Despite its remarkable success on relatively small astronomical scales, such as
the Solar System and compact astrophysical objects, Einstein's General
Relativity (GR) presently faces two deep conceptual crisis, related to the dark
energy and the dark matter problem. The dark energy problem was raised by
several high precision astronomical observations of the distant Type Ia
Supernovae, which have provided the unexpected result that in the Universe a
transition to an accelerating, de Sitter type phase has taken place recently
\cite{1n,2n,3n,4n,acc}. An equally intriguing question is related to the
matter-energy balance of the Universe. In order to close it according to the
cosmological observations, a second, and equally mysterious component of the
Universe, called Dark Matter, is necessary. Dark Matter is usually assumed
to be a non-baryonic and non-relativistic (cold) component of the Universe.
Its introduction is necessary on a fundamental level for explaining the
observed dynamics of the hydrogen clouds rotating around galaxies, which
have flat, non-decaying rotation curves, as opposed to the expected
Keplerian velocities. A second observation requiring the presence of dark
matter is the virial mass discrepancy in clusters of galaxies \cite{dm1,dm2}.
Up to now, no direct detection/observation of the dark matter has been
reported, and presently the only evidence for its  existence is its  gravitational interaction with baryonic matter.
Presently, after a long period of intensive observational and experimental
efforts the particle nature of the dark matter is still unknown.

Hence, these astronomical observations strongly suggest that at large scales
the force of gravity may not behave according to standard GR, as derived from the Hilbert-Einstein action, $S=\int{%
\left(R/2\kappa ^2+L_m\right)\sqrt{-g}d^4x}$, where $R$ is the Ricci scalar,
$\kappa $ is the gravitational coupling constant, and $L_m$ is the matter
Lagrangian, respectively, and that a generalization of the Hilbert-Einstein
action may be required for a full understanding of the gravitational interaction.
One of the promising ways to extend GR is related
to the modification of the geometric part of the Hilbert-Einstein
Lagrangian. Such an approach was introduced in \cite{Bu,Ba} by assuming that
the geometric part of the action is given by an arbitrary function $f(R)$ of
the Ricci scalar, so that the total Hilbert-Einstein action can be written
as $S=\int{\left(f(R)/2\kappa ^2+L_m\right)\sqrt{-g}d^4x}$. For in depth
discussions and reviews of modified and $f(R)$ type gravity theories see
\cite{r1,r2,r3,r4,rn, r5,r6,r7,r8}. The $f(R)$ modified theories of gravity can give a
satisfactory explanation to the recent cosmological observations, and it can
also provide a solution to the dark matter problem, which can be interpreted
as a geometric effect in the framework of the theory \cite{dm3}.

It is well known that Einstein's GR can be derived in two
different theoretical frameworks, the metric and the Palatini formalisms
\cite{Olmo}, respectively. Once applied to the Hilbert-Einstein action these
two approaches lead to the same equations of motion. However, this is not
the case in $f(R)$ gravity, and for other extended theories of
gravity, where it turns out that the field equations obtained using the
metric approach are generically different from their Palatini (or metric-affine) counterparts \cite{Olmo}. While the metric approach
typically leads to higher-order derivative field equations, in the Palatini
approach the resulting equations of motion are always second-order. However,
in the Palatini formulation certain algebraic relations between the matter
fields and the affine connection appear, with
the latter being now determined by a set of equations coupling it to the
matter fields and the metric. An extension of the $f(R)$
gravity theory, based on a hybrid combination of the metric and Palatini
mathematical formalisms, in which the (purely metric) Einstein-Hilbert
action is supplemented with (metric-affine) correction terms
constructed a la Palatini, was proposed in \cite{h1}. Both the
metric and the Palatini $f(R)$ theories allow the formulation of simple
extensions of GR with interesting properties.
However, at the same time, they each suffer from different types of
pathologies. Therefore establishing a bridge between these two apparently
different approaches may offer a possibility of eliminating their individual
pathologies. Further generalizations of the $f(R)$ gravity theories involving a geometry-matter coupling were proposed in \cite{Bert} and \cite{Harkof}, respectively.

Hence, in \cite{h1,h2} a hybrid combination of the
metric and Palatini formalisms was used to construct a gravitational
Lagrangian. As a main result of this approach it was found that viable
models containing elements of both formalisms are possible. An important
result of these theories is the possibility to generate long-range forces
without entering into conflict with the local Solar System
tests of gravity. An important technical result is the possibility of using
a scalar-tensor representation for the hybrid metric-Palatini theories,
which simplifies the analysis of the field
equations and the construction of solutions. An example of such hybrid
metric-Palatini theory is the one based on the gravitational Lagrangian $R +
f(\mathcal{R})$, where $\mathcal{R}$ is the Palatini scalar curvature. By
introducing such an action means that we maintain all the positive results
of GR at the scale of the Solar System and of compact
objects, which are included in the Einstein-Hilbert part of the action $R$,
while the metric-affine $f(\mathcal{R})$ component adds novel
features that could explain the recent cosmological observations. A related
formalism for the study of $f(R)$ theories that interpolate between the
metric and Palatini regimes, and called C-theory, was proposed in \cite%
{Ko1,Ko2}. A generalization of the hybrid metric-Palatini gravity was
proposed in \cite{B1}.

Much attention has been invested in the hybrid metric-Palatini gravity. In a cosmological context, the properties of the Einstein static Universe were studied in \cite{B}. The cosmological applications of metric-Palatini gravity were explored in \cite{c1}, and cosmological solutions coming from the scalar-tensor representation were presented. Criteria to obtain the late-time cosmic acceleration were discussed, and the field equations were analyzed as a dynamical system. Several classes of dynamical cosmological solutions, depending on the functional form of the effective scalar field potential, describing both accelerating and decelerating Universes were also explicitly obtained.
The evolution of the linear perturbations in the hybrid metric-Palatini
theory was studied in \cite{c4}, where the full set of linearized evolution
equations for the perturbed potentials were derived. It turns out that the
main deviations from the $\Lambda$CDM model arise in the distant past, with an oscillatory signature in the ratio between the Newtonian potentials $\Phi $ and $\Psi$.
Two classes of models were studied in \cite{c6}, where both models recover GR with an effective Cosmological Constant at late times. This occurs because the Palatini Ricci scalar evolves towards and asymptotically
settles at the minimum of its effective potential during the cosmological
evolution. With the use of a combination of cosmic microwave background,
supernovae and baryonic acoustic oscillations the free parameters of the models were constrained. It is interesting to note that for both models considered, the maximum deviation from the gravitational constant $G$ is of the order of 1\%.
The cosmology of the metric-Palatini theories was also studied using the dynamical system approach in \cite{c5} by formulating the propagation equation as an autonomous system. The analysis resulted in the standard cosmological fixed points and new accelerating solutions were found that can be attractors in the phase space.

In the context of dark matter, the virial theorem for galaxy clusters in hybrid metric-Palatini gravity was derived in \cite{c2}, where it was shown that the total virial mass is proportional to the effective mass associated with the new terms generated by the effective scalar field, and the baryonic mass. Hence, the geometric
terms in the generalized virial theorem may account for the virial mass
discrepancy in clusters of galaxies. Astrophysical applications of the model
were also considered, and it was shown that the model predicts that the mass
associated to the scalar field and its effects extend beyond the virial
radius of the clusters of galaxies.  The possibility that the behavior of
the rotational velocities of test particles gravitating around galaxies can
be explained within the framework of the hybrid metric-Palatini
gravitational theory was investigated in \cite{c3}. The tangential velocity
of test particles can be explicitly obtained as a function of the scalar
field of the equivalent scalar-tensor description. Therefore, all the physical
and geometrical quantities and the numerical parameters in the hybrid
metric-Palatini model can be expressed in terms of observable/measurable
parameters, such as the tangential velocity, the baryonic mass of the
galaxy, the Doppler frequency shifts, and the stellar dispersion velocity,
respectively. Furthermore, the well-formulation and well-posedness of the Cauchy problem was discussed for hybrid metric-Palatini gravity in \cite{P}. Wormhole solutions have also been obtained in the hybrid metric-Palatini theory \cite{Lobo}, where the higher order terms support theses exotic geometries.
For a recent review of hybrid metric-Palatini gravity see \cite{rev}.

Spherical symmetry has played an important role in GR, since
a large class of solutions of Einstein's gravitational field equations,
describing the interior structure of relativistic compact objects, can be
obtained under this assumption. 
The search for exact solutions describing static neutral,
charged, isotropic or anisotropic stellar type configurations has
continuously attracted the interests of the scientific community. A huge number of
analytical solutions of the Einstein gravitational field equations describing the interior structure of the
static fluid spheres were found in the past 100 years (for reviews of the
interior solutions of the Einstein gravitational field equations  see \cite{0,1b,2b}).
The study of the stellar structure can also provide important constraints on
modified theories of gravity. Presently, a large number of neutron star
masses are available, due to a significant increase in the precision of the
observations \cite{Oz,Hor}. These observations have revealed an intrinsically
complex distribution of the masses of the neutron stars, with the important
conclusion that last century's paradigm that there is single, $1.4M_{\odot}$
mass scale, is not supported by the astronomical data. A bimodal or even
more complex distribution can actually be seen in the numerical data \cite%
{Hor}. Observations performed  through pulsar timing \cite{Demorest, Antoniadis},  have confirmed with a high precision that some neutron stars have
masses of around $2M_{\odot}$. On the other hand firm limits on the maximum and minimum
values of the neutron star masses in nature are still unknown. Besides the
information of the maximum masses and radii of neutron or other stars,
observations of the surface gravitational redshift can also provide
important constraints on modified theories of gravity. The structure and physical properties of specific classes of neutron, quark
and exotic stars in various modified gravity theories have been extensively studied in \cite{s1,s2,s3,s4,s5,s6,s7,s8,s9,s10,s11,s12}.

It is the goal of this paper to investigate the properties of relativistic
compact high density stars in the hybrid metric-Palatini theory in its
scalar-tensor version. By adopting a spherically symmetric geometry and a
perfect fluid matter source, as a first step in our study we
obtain the mass continuity equation and the Tolman-Oppenheimer-Volkoff
equation, describing, together with the generalized Klein-Gordon equation
satisfied by the scalar field, the macroscopic properties of the star. The structure equations of the hybrid metric-Palatini theory are then solved
numerically for several prescribed equations of state of the dense matter.
As specific examples of high density compact objects, we
consider stars described by the causal stiff fluid
(Zeldovich) equation of state, with the property that the speed of sound in
the dense matter equals the speed of light; the radiation-type equation of
state, describing a photon gas, for which the trace of the energy-momentum
tensor is zero; the quark matter equation of state, and, finally, the
Bose-Einstein Condensate equation of state, corresponding to a polytropic
equation of state with polytropic index $n=1$. For all these physical models
the global astrophysical parameters of the stars (radius and mass), as well
as the scalar field, are obtained in both standard GR and in
the hybrid metric-Palatini gravity theory. This procedure allows an in depth
comparison of the two approaches for the description of stellar structure
and properties. As a general conclusion of our study we find that hybrid
metric-Palatini gravity allows the existence of more massive stars, as
compared to GR. Furthermore, two classes of
hybrid metric-Palatini stellar models, corresponding to two fixed forms of the scalar field, are also
investigated in detail. An interesting result of this analysis is that in the case of a constant scalar field, which is the minimum of a Higgs type potential,  the equation of state of the matter takes the form of the bag model equation
of state, describing quark matter.

The present paper is organized as follows. The hybrid metric-Palatini gravity theory is briefly presented in Section~\ref{sect2}. The system of gravitational field equations, describing the star interior, are presented in Section~\ref{sect3}, where the structure equations of the star (mass continuity, Tolman-Oppenheimer-Volkoff, and Klein-Gordon) are also derived, and reformulated in a dimensionless form. The structure and global astrophysical parameters of stiff fluid, radiation fluid, quark matter and Bose-Einstein Condensate stars are obtained, by numerically integrating the structure equations, in Section~\ref{sect5}. Stellar models with fixed forms of the scalar field are analyzed in Section~\ref{sect4}. We discuss and conclude our results in Section~\ref{sect6}.

\section{Hybrid metric-Palatini gravity: Formalism}\label{sect2}

The action for the hybrid metric-Palatini gravity is \cite{h1}
\begin{equation}  \label{eq:S_hybrid}
S=\frac{1}{2\kappa^2}\int d^4 x \sqrt{-g} \left[ R + f(\mathcal{R})\right] +
S_m,
\end{equation}
where $\kappa^2\equiv 8\pi G_0/c^4$, with $G_0$ and $c$ denoting the standard gravitational constant, and the speed of light, respectively, $S_m$ is the matter action, $R$ is the
metric Einstein-Hilbert term, $\mathcal{R} \equiv g^{\mu\nu}\mathcal{R}%
_{\mu\nu} $ is the Palatini curvature, and $\mathcal{R}_{\mu\nu}$ is defined
in terms of an independent connection $\hat{\Gamma}^\alpha_{\mu\nu}$ as
\begin{equation}
\mathcal{R}_{\mu\nu} \equiv \hat{\Gamma}^\alpha_{\mu\nu ,\alpha} - \hat{%
\Gamma}^\alpha_{\mu\alpha , \nu} + \hat{\Gamma}^\alpha_{\alpha\lambda}\hat{%
\Gamma}^\lambda_{\mu\nu} -\hat{\Gamma}^\alpha_{\mu\lambda}\hat{\Gamma}%
^\lambda_{\alpha\nu}\,.
\end{equation}

Varying the action (\ref{eq:S_hybrid}) with respect to the metric, one
obtains the following gravitational field equations
\begin{equation}  \label{efe}
G_{\mu\nu} + F(\mathcal{R})\mathcal{R}_{\mu\nu}-\frac{1}{2}f(\mathcal{R}%
)g_{\mu\nu} = \kappa^2 T_{\mu\nu}\,,
\end{equation}
where the matter energy-momentum tensor is defined as
\be
T_{\mu\nu} \equiv
-\left(\frac{2}{\sqrt{-g}}\right)\frac{ \delta (\sqrt{-g}\mathcal{L}_m)}{\delta g^{\mu\nu}}.
\ee
 The independent connection is compatible with the metric $F(\mathcal{R}%
)g_{\mu\nu}$, conformal to $g_{\mu\nu}$; the conformal factor is given by $F(%
\mathcal{R}) \equiv df(\mathcal{R})/d\mathcal{R}$. The latter considerations
imply that
\begin{eqnarray}  \label{ricci}
\mathcal{R}_{\mu\nu}  =  R_{\mu\nu} + \frac{3}{2}\frac{1}{F^2(\mathcal{R})}%
F(\mathcal{R})_{,\mu}F(\mathcal{R})_{,\nu}
- \frac{1}{F(\mathcal{R})}\nabla_\mu F(\mathcal{R})_{,\nu} - \frac{1}{2}%
\frac{1}{F(\mathcal{R})}g_{\mu\nu}\Box F(\mathcal{R})\,.
\end{eqnarray}
Note that $\mathcal{R}$ can be obtained from the trace of the field
equations (\ref{efe}), which yields: $F(\mathcal{R})\mathcal{R} -2f(\mathcal{R%
}) - R = \kappa^2 T$.

The hybrid metric-Palatini action (\ref{eq:S_hybrid}) can be turned into a scalar-tensor theory, by introducing an auxiliary field $E$, given by the
following action (we refer the reader to \cite{h1} for more
details)
\begin{equation}
S=\frac{1}{2\kappa^{2}}\int\mathrm{d}^{4}x\sqrt{-g}[R+f(E)+f^{\prime }(E)(%
\mathcal{R}-E)].
\end{equation}
The field $E$ is dynamically equivalent to the Palatini scalar $\mathcal{R}$
if $f^{\prime\prime}(\mathcal{R})\neq0$. Defining$\;$%
\begin{equation}
\phi\equiv f^{\prime}(E), \qquad V(\phi)=Ef^{\prime}(E)-f(E),   \label{lagr}
\end{equation}
the action becomes
\begin{equation}  \label{eq:S_scalar1}
S= \frac{1}{2\kappa^2}\int d^4 x \sqrt{-g} \left[ R + \phi\mathcal{R}-V(\phi)%
\right] +S_m \ .
\end{equation}
Varying this action with respect to the metric, the scalar $\phi$ and the
connection yields the following field equations
\begin{eqnarray}
R_{\mu\nu}+\phi \mathcal{R}_{\mu\nu}-\frac{1}{2}\left(R+\phi\mathcal{R}%
-V\right)g_{\mu\nu}&=&\kappa^2 T_{\mu\nu} \,,  \label{eq:var-gab} \\
\mathcal{R}-V_\phi&=&0 \,,  \label{eq:var-phi} \\
\hat{\nabla}_\alpha\left(\sqrt{-g}\phi g^{\mu\nu}\right)&=&0 \,, \
\label{eq:connection}
\end{eqnarray}
respectively.

It is useful to note that Eq. (\ref{lagr}) is a Clairaut differential equation \cite{Claraut}, that is,
\begin{equation}
Ef^{\prime }(E)-f(E)=V\left( f^{\prime }\left( E\right) \right) \,.
\label{fe.04}
\end{equation}%
It admits a general linear solution%
\begin{equation}
f\left( E\right) = h\, E-V\left( h\right) \,,  \label{fe.05}
\end{equation}
for arbitrary $V\left( \phi \right) $ and a singular solution followed from
the equation
\begin{equation}
\frac{\partial V\left( f^{\prime }\left( E\right) \right) }{\partial
f^{\prime }}-E=0 \,.  \label{fe.06}
\end{equation}

Note that the solution of Eq.~(\ref{eq:connection}) implies
that the independent connection is the Levi-Civita connection of a metric $%
h_{\mu\nu}=\phi g_{\mu\nu}$. Thus we are dealing with a bi-metric theory and
$\mathcal{R}_{\mu\nu}$ and $R_{\mu\nu}$ are related by
\begin{equation}  \label{eq:conformal_Rmn}
\mathcal{R}_{\mu\nu}=R_{\mu\nu}+\frac{3}{2\phi^2}\partial_\mu \phi
\partial_\nu \phi-\frac{1}{\phi}\left(\nabla_\mu \nabla_\nu \phi+\frac{1}{2}%
g_{\mu\nu}\Box\phi\right) \ ,
\end{equation}
and consequently 
\be
\mathcal{R}=R+\frac{3}{2\phi^2}\partial_\mu \phi \partial^\mu \phi-\frac{3}{%
\phi}\Box \phi,
\end{equation}
which can be used in the action (\ref{eq:S_scalar1}) to get rid of the
independent connection and obtain the following scalar-tensor representation
\cite{h1}
\begin{eqnarray}  \label{eq:S_scalar2}
S= \frac{1}{2\kappa^2}\int d^4 x \sqrt{-g} \left[ (1+\phi)R +\frac{3}{2\phi}%
\partial_\mu \phi \partial^\mu \phi -V(\phi)\right] +S_m .  \nonumber
\end{eqnarray}
It is important to note that this action differs fundamentally from the $%
w=-3/2$ Brans-Dicke theory in the coupling of the scalar to the curvature.

Now substituting Eq.~(\ref{eq:var-phi}) and Eq.~(\ref{eq:conformal_Rmn}) in
Eq.~(\ref{eq:var-gab}), the metric field equation can be written as an
effective Einstein field equation, i.e., $G_{\mu\nu}=\kappa^2 T^{\mathrm{eff}%
}_{\mu\nu}$, where the effective energy-momentum tensor is given by
\begin{eqnarray}
T^{\mathrm{eff}}_{\mu\nu}=\frac{1}{1+\phi} \left\{ T_{\mu\nu} - \frac{1}{%
\kappa^2} \left[ \frac{1}{2}g_{\mu\nu}\left(V+2\Box\phi\right)+
\nabla_\mu\nabla_\nu\phi-\frac{3}{2\phi}\partial_\mu \phi \;\partial_\nu
\phi + \frac{3}{4\phi}g_{\mu\nu}(\partial \phi)^2 \right] \ \right\} .
\label{effSET}
\end{eqnarray}

The scalar field is governed by the second-order evolution equation (we
refer the reader to \cite{h1} for more details)
\begin{equation}  \label{eq:evol-phi}
-\Box\phi+\frac{1}{2\phi}\partial_\mu \phi \partial^\mu \phi+\frac{\phi[%
2V-(1+\phi)V_\phi]} {3}=\frac{\phi\kappa^2}{3}T\,,
\end{equation}
which is an effective Klein-Gordon equation. This last expression shows
that, unlike in the Palatini ($w=-3/2$) case, the scalar field is dynamical.
Thus, the theory is not affected by the microscopic instabilities that arise
in Palatini models with infrared corrections \cite{Olmo}. As for the
matter energy-momentum tensor, it is conserved independently, so that $\nabla _{\mu }T^{\mu}_{\nu}=0$.

It is important to analyse the post-Newtonian parameters of the theory, in order to determine the viability of the theory with local gravitational tests. To this effect, we consider the post-Newtonian analysis and consider the perturbations of Eqs.  (\ref{effSET}) and (\ref{eq:evol-phi}) in a Minkowskian background. Consider $\phi=\phi_0+\varphi(x)$, where $\phi_0$ is the asymptotic value of the field far away from the local system, and a quasi-Minkowskian coordinate system in which $g_{\mu\nu}\approx \eta_{\mu\nu}+h_{\mu\nu}$, with $|h_{\mu\nu}|\ll 1$. This provides the standard post-Newtonian metric up to second order for this class of theories, with the following results (we refer the reader to Ref. \cite{rev} for details)
\begin{eqnarray}
G_{\rm eff}&\equiv & \frac{\kappa^2}{8\pi (1+\phi_0)}\left(1+\frac{\phi_0}{3}e^{-m_\varphi r}\right), \\
\gamma &\equiv & \frac{\left[1+\phi_0\exp \left(-m_{\varphi} r\right)/3\right]}{\left[1-\phi_ 0\exp \left(-m_{\varphi } r\right)/3\right]},\label{gamma0} \\
m_\varphi^2 &\equiv &\frac{1}{3} \left[
2V-V_{\phi}-\phi(1+\phi)V_{\phi\phi}\right]\big|_{\phi=\phi_0}\,. \label{mass}
\end{eqnarray}
In the hybrid metric-Palatini theory there are two possibilities to obtain that the PPN parameter is $\gamma\approx 1$. Note that the first one is the same as in $f(R)$ theories and involves a very massive scalar field \cite{rn}. The second possibility resides imposing $\phi_0\ll 1$, so that the Yukawa-type corrections are very small regardless of the magnitude of $m_\varphi$. This latter case could allow for the existence of a long-range scalar field able to modify the cosmological dynamics, but leaves the locat gravity tests unaffected.

\section{The hydrostatic equilibrium equations for spherically symmetric stars in hybrid metric-Palatini gravity}\label{sect3}

Consider the following line element in curvature coordinates, which
represents a static and spherically symmetric geometry
\begin{eqnarray}
ds^2=-e^{\nu(r)}c^2dt^2 + e^{\lambda(r)} dr^2 + r^2 \left(d \theta^2 + \sin
^2\theta d\varphi^2 \right) \,, \label{whmetric}
\end{eqnarray}
where the metric functions $\nu(r)$ and $\lambda(r)$ are functions of the
radial coordinate, and denoted the mass and the redshift functions,
respectively; with radial coordinate range $0 \leq r < \infty$. It is
possible to construct asymptotically flat spacetimes, in which $\nu(r)
\rightarrow 0$ and $\lambda(r) \rightarrow 0$ as $r \rightarrow \infty$. For
the matter energy-momentum tensor, $T_{\mu \nu}$, we adopt the perfect fluid
form, so that in the comoving frame with four-velocity $u^{\mu}=\left(e^{-%
\nu /2},0,0,0\right)$ it has the components $T^{\mu}_{\nu}=\mathrm{diag}%
\left(-\rho c^2, p_r,p_t,p_t\right)$, where $\rho$ is the energy density, $p_r$ and $p_t$ are the radial and tangential pressures, respectively.

Using the metric (\ref{whmetric}), the effective Einstein field equation (\ref{effSET}) provides the following gravitational field equations
\begin{eqnarray}
\kappa ^{2}\rho (r)c^{2} &=&\frac{1}{r^{2}}\left[ 1-e^{-\lambda }\left(
1-r\lambda ^{\prime }\right) \right] (1+\phi )-e^{-\lambda }\left[ \phi
^{\prime \prime }-\frac{3\phi ^{\prime 2}}{4\phi }\right]  \nonumber
\label{hybrid_rho} \\
&&+\frac{\phi ^{\prime }}{2r}e^{-\lambda }\left( r\lambda ^{\prime
}-4\right) -\frac{V}{2}\,,
\end{eqnarray}%
\begin{equation}
\kappa ^{2}p_{r}(r)=\left[ \frac{1}{r^{2}}(e^{-\lambda }-1)+\frac{\nu
^{\prime }}{r}e^{-\lambda }\right] (1+\phi )+\phi ^{\prime }\left( \frac{\nu
^{\prime }}{2}+\frac{2}{r}+\frac{3\phi ^{\prime }}{4\phi }\right)
e^{-\lambda }+\frac{V}{2}\,  \label{hybrid_pr}
\end{equation}%
\begin{eqnarray}
\kappa ^{2}p_{t}(r) &=&\Bigg[\left( \frac{\nu ^{\prime \prime }}{2}+\left(
\frac{\nu ^{\prime }}{2}\right) ^{2}+\frac{\nu ^{\prime }}{2r}\right)
e^{-\lambda }-\frac{1}{2}\frac{\lambda ^{\prime }e^{-\lambda }}{r}\left( 1+r%
\frac{\nu ^{\prime }}{2}\right) \Bigg](1+\phi )  \nonumber \\
&&+\left[ \phi ^{\prime \prime }+\frac{\phi ^{\prime }\nu ^{\prime }}{2}+%
\frac{3\phi ^{\prime 2}}{4\phi }\right] e^{-\lambda }+\frac{\phi ^{\prime }}{%
r}e^{-\lambda }\left( 1-\frac{r\lambda ^{\prime }}{2}\right) +\frac{V}{2}\,,
\end{eqnarray}
where we have denoted by a prime the derivative with respect to radial coordinate $r$.
The effective Klein-Gordon equation (\ref{eq:evol-phi}) is given by
\begin{equation}
-\left[ \phi ^{\prime \prime }+\frac{\phi ^{\prime }\nu ^{\prime }}{2}-\frac{%
\phi ^{\prime 2}}{2\phi }+\frac{2\phi ^{\prime }}{r}\right] e^{-\lambda }+%
\frac{\phi ^{\prime }\lambda ^{\prime }}{2}e^{-\lambda }+\frac{\phi }{3}%
\left[ 2V-(1+\phi )V_{\phi }\right] =\frac{\phi \kappa ^{2}}{3}T\,.
\label{modKGeq}
\end{equation}
The conservation of the matter energy-momentum tensor gives the following
relation between the components of the energy-momentum tensor, and the
metric tensor component $\nu $,
\begin{equation}
\nu ^{\prime }=-\frac{2p_{r}^{\prime }}{\rho c^2+p_{r}}+\frac{2\left(
p_{t}-p_{r}\right) }{r}.  \label{eqcons}
\end{equation}

Note that Eqs.~(\ref{hybrid_rho})-(\ref{modKGeq}) provide four independent
equations, for seven unknown quantities, i.e. $\rho (r)$, $p_{r}(r)$, $%
p_{t}(r)$, $\nu (r)$, $\lambda (r)$, $\phi (r)$ and $V(r)$. Thus, the system
of equations is underdetermined, so that we will reduce the number of
unknown functions by assuming suitable conditions. In the following we will
restrict our analysis to the isotropic pressure distribution case only, by
assuming $p_{r}=p_{t}=p$.

\subsection{The mass continuity and the Tolman-Oppenheimer-Volkoff equation}

As a first step in our analysis we divide Eq.~(\ref{hybrid_rho}) by $1+\phi $,
and introduce the effective gravitational coupling defined as $%
G_{eff}=G_{0}/(1+\phi )$, where $G_{0}$ is the standard general relativistic
gravitational constant. Hence we can introduce the effective gravitational
coupling denoted as $\kappa _{eff}^{2}=8\pi G_{eff}/c^4$. By taking into account
the mathematical identity
\begin{equation}
\phi ^{\prime \prime }-\frac{3\phi ^{\prime 2}}{4\phi }=\phi ^{3/4}%
\frac{d}{dr}\frac{\phi ^{\prime }}{\phi ^{3/4}}=4\phi ^{3/4}\frac{d^{2}}{%
dr^{2}}\phi ^{1/4},
\end{equation}
and by denoting $\phi =e^{\Phi }-1$ and $G_{eff}=G_{0}e^{-\Phi }$, so that $\phi ^{\prime }/(1+\phi)=\Phi ^{\prime }$, one arrives at the following relations
\begin{eqnarray}
\frac{1}{1+\phi }\left[ \phi ^{\prime \prime }-\frac{%
3\phi ^{\prime 2}}{4\phi }\right] &=& \frac{1}{4}\left( 1+\frac{3}{1-e^{\Phi }}%
\right) \Phi ^{\prime 2}+\Phi ^{\prime \prime }=f(\Phi ),  \nonumber \\
V(\phi )&=&\left( 1+\phi
\right) U(\Phi )=e^{\Phi }U\left( \Phi \right) \,. \label{potentialV}
\end{eqnarray}%
Equation~(\ref{hybrid_rho}) can be written as
\begin{equation}
\frac{d}{dr}re^{-\lambda }=-\frac{rf(\Phi )+3\Phi ^{\prime }/2}{1+\Phi
^{\prime }r/2}re^{-\lambda }+\frac{1-\kappa _{eff}^{2}\rho
c^{2}r^{2}-Ur^{2}/2}{1+\Phi ^{\prime }r/2}\mathbf{.}  \label{e1}
\end{equation}

By representing the metric tensor coefficient $e^{-\lambda }$\ as
\begin{equation}
e^{-\lambda }=1-\frac{2G_{0}m_{eff}(r)}{c^{2}r},
\end{equation}%
it follows that the effective mass $m_{eff}(r)$\ satisfies the differential
equation
\begin{equation}
\frac{dm_{eff}}{dr}=-\frac{rf(\Phi )+3\Phi ^{\prime }/2}{1+\Phi ^{\prime }r/2%
}m_{eff}+\frac{4\pi r^{2}}{\kappa ^{2}c^{2}\left[ 1+\Phi ^{\prime }r/2\right]
}\left[ 2\frac{\Phi ^{\prime }}{r}+\frac{U}{2}+f(\Phi )+\kappa
_{eff}^{2}\rho c^{2}\right] ,  \label{meq}
\end{equation}%
with the general solution given by
\begin{eqnarray}
m_{eff}(r) &=&\frac{4\pi }{\kappa ^{2}c^{2}}\exp \left[ -\int_{0}^{r}\frac{%
r^{\prime }f(\Phi \left( r^{\prime }\right) )+3\Phi ^{\prime }\left(
r^{\prime }\right) /2}{1+\Phi ^{\prime }\left( r^{\prime }\right) r^{\prime
}/2}dr^{\prime }\right] \int_{0}^{r}\exp \left[ \int_{0}^{r^{\prime }}\frac{%
r^{\prime \prime }f(\Phi \left( r^{\prime \prime }\right) )+3\Phi ^{\prime
}\left( r^{\prime \prime }\right) /2}{1+\Phi ^{\prime }\left( r^{\prime
\prime }\right) r^{\prime \prime }/2}dr^{\prime \prime }\right] \times
\nonumber \\
&&\times \frac{r^{\prime 2}}{\left[ 1+\Phi ^{\prime }\left( r^{\prime }\right)
r^{\prime }/2\right] }\left[ \frac{2\Phi ^{\prime }\left( r^{\prime }\right)
}{r^{\prime }}+\frac{U\left( \Phi \left( r^{\prime }\right) \right) }{2}%
+f(\Phi \left( r^{\prime }\right) )+\kappa _{eff}^{2}\rho \left( r^{\prime
}\right) c^{2}\right] dr^{\prime },
\end{eqnarray}
where we have used the transformation $c^{2}/2G_{0}=4\pi /\kappa ^{2}c^{2}$.
Equivalently, Eq.~(\ref{meq}) can be written as
\begin{equation}\label{26}
\frac{dm_{eff}}{dr}=4\pi \rho _{eff}r^{2},
\end{equation}%
where we have introduced the effective density of the star, defined as
\begin{equation}
\rho _{eff}=-\frac{rf(\Phi )+3\Phi ^{\prime }/2}{4\pi r^{2}\left[ 1+\Phi
^{\prime }r/2\right] }m_{eff}+\frac{1}{\kappa ^{2}c^{2}\left[ 1+\Phi
^{\prime }r/2\right] }\left[ 2\frac{\Phi ^{\prime }}{r}+\frac{U\left( \Phi
\right) }{2}+f(\Phi )+\kappa _{eff}^{2}\rho c^{2}\right] .
\end{equation}

Equation~(\ref{hybrid_pr}) can be solved for $\nu ^{\prime }$ to give
\begin{equation}
\mathbf{\nu }^{\prime }=\frac{\left( \kappa ^{2}pe^{-\Phi }-U/2\right)
r^{2}-\left( 1-2G_{0}m_{eff}/c^{2}r\right) \left\{ 1+r\left[ 2+rh(\Phi )%
\right] \Phi ^{\prime }\right\} +1}{r\left( 1-2G_{0}m_{eff}/c^{2}r\right)
\left( 1+\Phi ^{\prime }r/2\right) } \,, \label{nuprime}
\end{equation}
where we have defined
\begin{equation}
h(\Phi )=\frac{3e^{\Phi }\Phi ^{\prime }}{4\left( e^{\Phi }-1\right) }.
\end{equation}

Then, with the use of the energy-momentum conservation equation (\ref{eqcons}), we obtain the generalized Tolman-Oppenheimer-Volkoff equation, describing
the hydrostatic equilibrium of compact astrophysical objects in
hybrid metric-Palatini gravity as
\begin{equation}
\frac{dp}{dr}=-\frac{\left( \rho c^{2}+p\right) \left\{ \left( \kappa
^{2}pe^{-\Phi }-U/2\right) r^{2}-\left( 1-2G_{0}m_{eff}/c^{2}r\right)
\left\{ 1+r\left[ 2+rh(\Phi )\right] \Phi ^{\prime }\right\} +1\right\} }{%
r\left( 1-2G_{0}m_{eff}/c^{2}r\right) \left( 2+\Phi ^{\prime }r\right) }.
\label{peq}
\end{equation}

Finally, in the new scalar field variable, the Klein-Gordon equation (\ref%
{modKGeq}) takes the form
\begin{eqnarray}\label{31}
-\Phi ^{\prime \prime }+\frac{1}{2}\frac{2-e^{\Phi }}{e^{\Phi }-1}\Phi
^{\prime 2}-\Phi ^{\prime }\left[ -\frac{p^{\prime }}{\rho c^{2}+p}+\frac{2}{%
r}-\frac{G_{0}}{c^{2}}\frac{4\pi \rho _{eff}r^{3}-m_{eff}}{r^{2}\left(
1-2G_{0}m_{eff}/c^{2}r\right) }\right]   \nonumber \\
+\frac{ e^{\Phi }-1 }{3\left( 1-2G_{0}m_{eff}/c^{2}r\right) }%
\left[ U\left( \Phi \right) -\frac{dU\left( \Phi \right) }{d\Phi }-\kappa
_{eff}^{2}T\right] =0,
\end{eqnarray}
where we have used the following relations
\begin{equation}
\lambda ^{\prime }=\frac{2G_{0}}{c^{2}}\frac{1}{e^{-\lambda }r^{2}}\left( r%
\frac{dm_{eff}}{dr}-m_{eff}\right) ,
\end{equation}%
and
\begin{equation}
\frac{dV}{d\phi }=\frac{d}{d\phi }\left[ U\left( 1+\phi \right) \right]
=U\left( \Phi \right) +\frac{dU\left( \Phi \right)
}{d\Phi },
\end{equation}%
respectively.

The system of equations  (\ref{meq}), (\ref{peq}) and (\ref{modKGeq}) must be
solved, after specifying an equation of state for the matter inside the
star, $p=p(\rho)$, with the boundary conditions $m_{eff}(0)=0$, $\rho
(0)=\rho _c$, $\Phi (0)=\Phi _0$, $\Phi ^{\prime }(0)=\Phi ^{\prime }_0 (0)$%
, and $p(R)=0$, respectively, where $\rho _c$ is the central density, and $R$
is the radius of the star, respectively. However, due to the singular nature of the center of the star, corresponding to the point $r=0$, when numerically integrating the gravitational field equations one must  impose the initial conditions at a small but nonzero radius $r=r_0$ \cite{DEF}, so that $m_{eff}\left(r_0\right)=0$, $\rho
\left(r_0\right)=\rho _c$ etc. On the other hand we must determine  the initial values of the radial derivatives of $\Phi $ at the center, $\Phi '\left(r_0\right)$ so that they are consistent with a regular
Taylor expansions at the origin, which can be given, for example, as \cite{DEF}
\be
\Phi (r)=\Phi\left(0\right)+\frac{1}{6}r^2\Delta \Phi (0)+{\rm O}\left(r^4\right),
\ee
where $\Delta \Phi (0)=\Phi (r)-\Phi (0)$. This series expansion determines the derivative of the scalar field as
\be
\lim _{r\rightarrow r_0} \Phi '(r)\approx \frac{1}{3}r_0\Delta \Phi (0).
\ee
Near the origin we can represent the effective mass as $m_{eff}(r)\sim 4\pi r^3\rho _c/3$. By taking into account the limits $\lim _{r\rightarrow r_0} \Phi '^{2}(r)= 0$, $\lim _{r\rightarrow r_0}p'(r)=0$, $\lim _{r\rightarrow r_0}m_{eff}/r=0$, $\lim _{r\rightarrow r_0}m_{eff}/r^2=0$, as well as the relation $\Phi ''(r)=\Delta \Phi (0)$, from Eq.~(\ref{31}) we obtain
\be
\Delta \Phi (0)=\frac{e^{\Phi_0}-1}{3}\left[U\left(\Phi_0\right)-\left.\frac{dU(\Phi)}{d\Phi}\right|_{\Phi =\Phi _0}+\frac{8\pi G}{c^4}e^{-\Phi _0}\left(\rho _cc^2-3p_c\right)\right],
\ee
giving for the central value of the derivative of the scalar field the expression
\be
\lim _{r\rightarrow r_0} \Phi '(r)=\frac{e^{\Phi_0}-1}{9}r_0\left[U\left(\Phi_0\right)-\left.\frac{dU(\Phi)}{d\Phi}\right|_{\Phi =\Phi _0}+\frac{8\pi G}{c^4}e^{-\Phi _0}\left(\rho _cc^2-3p_c\right)\right].
\ee

\subsection{Dimensionless form of the mass continuity, Tolman-Oppenheimer-Volkoff and Klein-Gordon equations}

In the following we will introduce a set of dimensionless variables $\left(
\eta ,\theta ,M_{eff},P,u\right) $, defined as
\begin{equation}  \label{dimvar}
r=a\eta , \qquad \rho =\rho _{c}\theta , \qquad  m_{eff}=M_{0}M_{eff}, \qquad
p=\rho_{c}c^{2}P, \qquad u=a^{2}U,
\end{equation}%
where
\begin{equation}\label{dima}
a=\frac{c}{\sqrt{8\pi G_{0}\rho _{c}}}, \qquad  M_{0}=\frac{ac^{2}}{G_{0}}=\frac{c^{3}
}{\sqrt{8\pi G_{0}^{3}\rho _{c}}}.
\end{equation}

In the new variables Eqs.~(\ref{modKGeq}), (\ref{meq}) and (\ref{peq}) take
the following dimensionless form
\begin{equation}
\frac{dM_{eff}}{d\eta }=-\frac{\eta f\left( \Phi \left( \eta \right) \right)
+(3/2)\left( d\Phi /d\eta \right) }{1+\eta \left( d\Phi /d\eta \right) /2}%
M_{eff}+\frac{\eta ^{2}}{2\left[ 1+\eta \left( d\Phi /d\eta \right) /2\right]
}\Bigg[\frac{2}{\eta }\frac{d\Phi }{d\eta }+\frac{u}{2}+f\left( \Phi (\eta
)\right) +\theta e^{-\Phi}\Bigg],  \label{dmeq}
\end{equation}
\begin{equation}
\frac{dP}{d\eta }=-\frac{\left( \theta +P\right) \left\{ \left( Pe^{-\Phi
}-u/2\right) \eta ^{2}-\left( 1-2M_{eff}/\eta \right) \left\{ 1+\eta \left[
2+\eta h(\Phi \left( \eta \right) )\right] \left( d\Phi /d\eta \right)
\right\} +1\right\} }{\eta \left( 1-2M_{eff}/\eta \right) \left[ 2+\eta
\left( d\Phi /d\eta \right) \right] },  \label{dpeq}
\end{equation}
\begin{eqnarray}  \label{dKGeq}
-\frac{d^{2}\Phi }{d\eta ^{2}}+\frac{1-e^{\Phi }/2}{e^{\Phi }-1}\left(
\frac{d\Phi }{d\eta }\right) ^{2}-\frac{d\Phi }{d\eta }\left[ -\frac{%
dP/d\eta }{\theta +P}+\frac{2}{\eta }-\frac{\rho _{eff}\left( \eta \right)
\eta ^{3}/2-M_{eff}}{\eta ^{2}\left( 1-2M_{eff}/\eta \right) }\right]
\nonumber \\
+\frac{\left( e^{\Phi }-1\right) }{3\left( 1-2M_{eff}/\eta \right) }\left\{ %
\left[ u\left( \Phi \right) -\frac{du\left( \Phi \right) }{d\Phi }\right]
-e^{-\Phi }\left( -\theta +3P\right) \right\} =0,
\end{eqnarray}
respectively, where
\begin{equation}
\rho _{eff}\left( \eta \right) =-2\frac{\eta f(\Phi \left( \eta \right)
)+3\left( d\Phi /d\eta \right) /2}{\eta ^{2}\left[ 1+\eta \left( d\Phi
/d\eta \right) /2\right] }M_{eff}+\frac{1}{1+\eta \left( d\Phi /d\eta
\right) /2} \Bigg[\frac{2}{\eta }\frac{d\Phi }{d\eta }+\frac{%
u\left(\Phi\right)}{2}+f\left( \Phi (\eta )\right) +\theta e^{-\Phi}\Bigg].
\end{equation}

The system of equations (\ref{dmeq})-(\ref{dKGeq}) must be integrated with
the boundary conditions $M_{eff}(0)=0$, $\theta (0)=1$, $\Phi (0)=\Phi _{0}$%
, $\left. \left( d\Phi /d\eta \right) \right\vert _{\eta =0}=\Phi
_{0}^{\prime }$, respectively, once the equation of state of the matter $%
P=P\left( \theta \right) $ has been chosen. As for the numerical value of the derivative of the scalar field at the center of the star it can be obtained in a dimensionless form as
\be
\lim _{\eta \rightarrow \eta_0} \Phi '(\eta )=\frac{e^{\Phi_0}-1}{9}\eta _0\left[u\left(\Phi_0\right)-\left.\frac{du(\Phi)}{d\Phi}\right|_{\Phi =\Phi _0}+e^{-\Phi _0}\left(1-3P_c\right)\right],
\ee
where $P_c$ is the value of the dimensionless pressure $P$ at the center of the star.

We will consider specific
numerical solutions describing the structure of the stars in hybrid
metric-Palatini gravity for a given equation of state of dense matter in the next Section.

\section{Structure of high density compact objects in hybrid metric-Palatini gravity}
\label{sect5}

In the present Section, we will investigate the properties of high density
stars in the hybrid metric-Palatini theory without imposing any
restrictions on the functional form of the scalar field $\Phi $. In the next Section we will investigate the field equations under the assumption that the scalar field $\Phi $ has a specific mathematical form, which is not determined dynamically by the field equations. In this latter case, after imposing the functional form of $\phi $, one can obtain from the field equations either the form of the equation of state of the matter, or the dynamical behavior of the scalar field potential associated to the a priori given form of the scalar field.

As for the
equation of state of the matter, we will consider four cases, corresponding
to the stiff fluid equation of state, with $P=\theta $, the radiation fluid
equation of state with $P=\theta /3$, the quark matter equation of state $%
P=\left(\theta -4b\right)/3$, respectively, and to the Bose-Einstein
Condensate superfluid neutron matter equation of state $P\propto \theta ^2$,
respectively.

In the following, we assume for all cases that the potential $U(\Phi)$ is of the Higgs type
\be
U(\Phi)=-\frac{\mu ^2}{2}\Phi ^2+\frac{\xi }{4}\Phi ^4,
\label{potentialU}
\ee
where $\mu ^2$ and $\xi $ are constants.
We also assume that similarly to the standard case, the constant $\mu^{2}<0$ is related to the mass of the hybrid metric-Palatini scalar particle by the relation $m_{\Phi}^2=2\xi v^2=-2\mu ^2$, where $
v^{2}=-\mu^{2}/\xi $ gives the minimum of the potential. For the case of strong interactions the Higgs
self-coupling constant $\lambda \approx 1/8$ \cite{Higgs}, which is a value inferred
on the determination of the mass of the Higgs boson from accelerator experiments. The dimensionless form $u(\Phi)$ of the potential is given by
\be
u(\Phi)=-\frac{\mu _0^2}{2}\Phi ^2+\frac{\xi_0 }{4}\Phi ^4,
\ee
where $\mu _0^2=a^2\mu ^2$, and $\xi _0=a^2\xi $, respectively, with $a^2$ given by Eq.~(\ref{dima}). It is important to note that in the dimensionless representation of the potential the coefficients $\mu _0^2$ and $\xi _0$ are functions of the central density of the star.

In all cases, we will compare our results with the standard general
relativistic spherically symmetric stellar models, described by the
structure equations \cite{Shap}
\begin{equation}\label{mgr1}
\frac{dm}{dr}=4\pi \rho r^2,
\end{equation}
\begin{equation}\label{pgr1}
\frac{dp(r)}{dr}=-\frac{\left(G_0/c^2\right)\left[\rho (r)c^2+p(r)\right]%
\left[\left(4\pi /c^2\right)p(r)r^3+m(r)\right]}{r^2\left(1-2G_0m(r)/c^2r\right)}.
\end{equation}
In the dimensionless variables introduced in Eqs.~(\ref{dimvar}), the
general relativistic structure equations take the dimensionless form
\begin{equation}\label{mgr}
\frac{dM_{GR}(\eta)}{d\eta }=\eta ^2\theta _{GR}(\eta),
\end{equation}
\begin{equation}\label{TOVgr}
\frac{dP_{GR}(\eta )}{d\eta }=-\frac{\left[\theta _{GR}(\eta)+P_{GR}(\eta)\right]\left[%
P_{GR}(\eta )\eta ^3+M_{GR}(\eta)\right]}{\eta ^2\left(1-2M_{GR}(\eta)/\eta\right)},
\end{equation}
respectively, which after imposing an equation of state $P=P(\theta)$ must be integrated
with the boundary conditions $\theta _{GR}(0)=1$ and $\theta _{GR}\left(\eta
_S\right)=0$, respectively.

\subsection{Explicit form of $f(\R)$}

Before proceeding with the analysis with the general relativistic spherically symmetric stellar models, it is interesting to arrive at a specific form for the $f(\R)$, by taking into account the potential (\ref{potentialU}). Now, using Eq. (\ref{eq:var-phi}), one arrives at
\begin{equation}  \label{14}
\mathcal{R}=\frac{dV(\phi)}{d\phi}=-\mu ^2 \ln (1+\phi ) \left\{1-\ln
(1+\phi ) \left[\frac{\xi }{\mu ^2} \ln (1+\phi ) \Bigg(\frac{1}{4}\ln
(1+\phi )+1\Bigg)-\frac{1}{2}\right]\right\}\,,
\end{equation}

Using this result, one cannot solve Eq. (\ref{lagr}) to provide the exact explicit form for $f(\mathcal{R})$. However, by performing a series expansion of Eq.~(\ref{14}) around the point $\phi =0$ [which is compatible with the analysis leading to Eqs. (\ref{gamma0})-(\ref{mass})], we obtain
\begin{equation}
\mathcal{R}=-\mu ^2 \phi + \left(\frac{\mu ^2}{6}+\xi \right)\phi ^3-\frac{5%
}{24} \left(\mu ^2+6 \xi \right)\phi ^4+O\left(\phi ^5\right).
\end{equation}
Similarly, for the potential $V(\phi )$ (\ref{potentialV}), we obtain
\begin{equation}
V(\phi )=-\frac{\mu ^{2}\phi ^{2}}{2}+\frac{1}{4}\left( \frac{\mu ^{2}}{6}%
+\xi \right) \phi ^{4}+O\left( \phi ^{5}\right) .
\end{equation}%
In the first order of approximation, $V(\phi )\approx -\mu ^{2}\phi ^{2}/2$,
and the Clairaut equation (\ref{lagr}) becomes
\begin{equation}
- \frac{\mu ^{2}}{2}\left[ f^{\prime }(E)\right] ^{2}=Ef^{\prime }(E)-f(E)
\end{equation}
(recall $E=\R$) which yields the general solution
\begin{equation}
f\left( E\right) =c_{1}E+\frac{c_{1}^{2}}{2}\mu ^{2}\,.
\end{equation}
where $c_{1}$ is an arbitrary constant of integration.
In the next order of
approximation we obtain for $f(E)$ the Clairaut type equation
\begin{equation}
-\frac{\mu ^{2}}{2}\left[ f^{\prime }(E)\right] ^{2}+\frac{1}{24}\left( \mu
^{2}+6\xi \right) \left[ f^{\prime }(E)\right] ^{4}=Ef^{\prime }(E)-f(E),
\end{equation}%
with the general solution
\begin{equation}
f(E)=c_{1}E+\frac{c_{1}^{2}}{2}\mu ^{2} -\frac{1}{4}\left( \frac{\mu ^{2}}{6}+\xi
\right) c_{1}^{4},
\end{equation}
where $c_{1}$ is an arbitrary integration constant.

By iteratively continuing this process we arrive at the conclusion that for the adopted
functional form of the potential we generally have
\begin{equation}
f(E)=c_{1}E+c_{2}(c_{1},\mu ,\xi ),
\end{equation}%
where the constant $c_{1}$ must be determined from some appropriate physical
requirements, or initial/boundary conditions.
It is interesting to note that this solution corresponds to a solution $f(\R)= \R$ with an effective cosmological constant given by $c_{2}(c_{1},\mu ,\xi)$.

On the other hand, in the present approach the numerical values of the constant $c_2$, which is essentially generated by the existence of the coupling,  are dependent on the  parameters $\left(\mu, \xi\right)$ of the adopted Higgs type potential. In order to study the effects of the coupling on the stellar structure   we will vary the values of the potential parameters in our numerical investigations. For each investigated stellar structure (corresponding to a specific equation of state of the dense matter) we also present the standard general relativistic result, which is obtained by suppressing the coupling by taking the limits $\mu \rightarrow 0$ and $\xi \rightarrow 0$, respectively. Hence our investigations can provide a clear picture of the effects of the variation of the gravitational couplings on the structure of high density compact astrophysical objects.

\subsection{Stiff fluid stars}

The equation of state $P=\theta $ is called the stiff (Zeldovich) equation
of state, and it gives the upper limit for the equation of state of a hot
nucleonic gas. It is believed that matter actually behaves in this manner at
densities above about ten times the nuclear density, that is, at densities
greater than $10^{17}\;\mathrm{g/cm^3}$, and at temperatures $T=\left(\rho
/\sigma \right)^{1/4}>10^{13}\;\mathrm{K}$, where $\sigma $ is the radiation
constant \cite{Shap}. For this equation of state the speed of sound is $c_s^2=\partial
P/\partial \theta =1$, so that the speed of matter perturbations cannot
exceed the speed of light. The stiff matter equation of state plays an
important role in astrophysics. By using the spherically symmetric static
Einstein field equations, the principle of causality, and Le Chatelier's
principle, it was shown in \cite{Ruf} that the maximum mass of the
equilibrium configuration of a high density neutron star cannot exceed the
upper limit of $3.2M_{\odot}$. To obtain this fundamental result it was
assumed that for high densities the equation of state of the neutron matter
is the stiff fluid equation of state $p=\rho c^2$. The numerical value of
the absolute maximum mass of a neutron star represents a fundamental method
for distinguishing observationally neutron or other compact stars from black
holes.

\begin{figure*}[h]
\centering
\includegraphics[width=9.5cm]{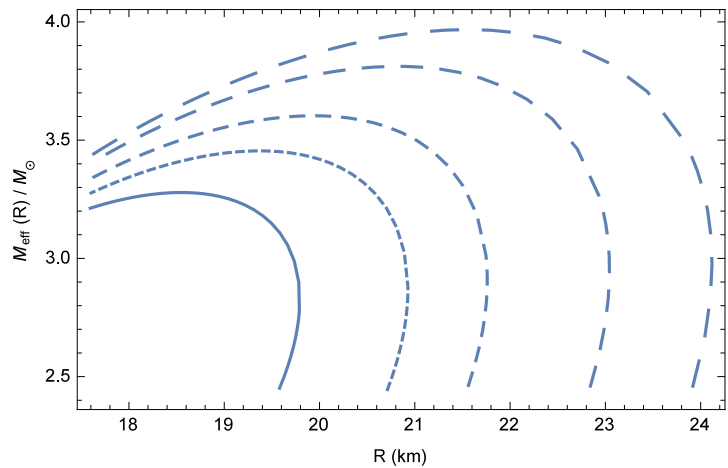}
\caption{Mass-radius relation for stiff fluid stars in hybrid metric-Palatini gravity theory, for $\mu = 10^{-5}\;{\rm cm ^{-1}}$, $\xi = 8.5\times 10^{-10}\;{\rm cm ^{-2}}$, $\Phi ' (0) = -1.8\times 10^{-16}\;{\rm cm}^{-1}$ to $-5.7 \times 10^{-16} \;{\rm cm^{-1}}$ for the central densities considered, and for different values of $\Phi (0)$: $\Phi \equiv 0$ (standard general relativistic limit) (solid curve), $\Phi (0) = 0.14$ (dotted curve),  $\Phi (0)=0.22$ (short dashed curve), $\Phi (0)=0.28$ (dashed curve), and $\Phi (0)= 0.30$ (long dashed curve).}
\label{f1_stiff}
\end{figure*}

The  mass-radius relations for stiff fluid stars in both standard general relativity and hybrid metric Palatini gravity theory are represented in Fig.~\ref{f1_stiff}.

The properties of this class of stars have been obtained by numerically integrating the star structure Eqs.~(\ref{meq}), (\ref{peq}) and (\ref{31}) for the stiff fluid equation of state. In order to obtain the plots we have chosen for the coefficients $\mu $ and $\xi $ in the Higgs potential the values $\mu = 10^{-5}\;{\rm cm ^{-1}}$, and $\xi = 8.5\times 10^{-10}\;{\rm cm ^{-2}}$, respectively. Then, we have varied the central values of the scalar field $\Phi (0)$, thus generating several sequences of stable stiff fluid stellar models. The derivative of the scalar field $\Phi '$ at the center of the star was calculated to be  between $\Phi ' (0) = -1.8\times 10^{-16} \;{\rm cm^{-1}}$ and $\Phi ' (0) = -5.7\times 10^{-16} \;{\rm cm^{-1}}$ in the central density interval considered. In order to compare the structure of the stars in hybrid metric-Palatini gravity and standard general relativity we also present the mass-radius relation for stiff fluid stars, obtained as the solutions of the general relativistic mass continuity and TOV equations (\ref{mgr1}) and (\ref{pgr1}), respectively.

The central density was modified  between the values $3.1\times 10^{14}\;{\rm g/cm^3}$, and $2.9\times 10^{15}\;{\rm g/cm^3}$ for the hybrid metric-Palatini gravity stars, with $\Phi \neq 0$,  and in the range $3.1\times 10^{14}\;{\rm g/cm^3}$ and $2.2\times 10^{15}\;{\rm g/cm^3}$ for the standard general relativistic model. The maximum masses for these stellar sequences are $M_{max}=3.278M_{\odot}$ (corresponding to the standard general relativistic maximum mass value \cite{Ruf}), $M_{max}=3.454M_{\odot}$, $M_{max}=3.603M_{\odot}$, $M_{max}=3.811M_{\odot}$ and $M_{max}=3.968M_{\odot}$, respectively.

\begin{figure*}[h]
\centering
\includegraphics[width=8.5cm]{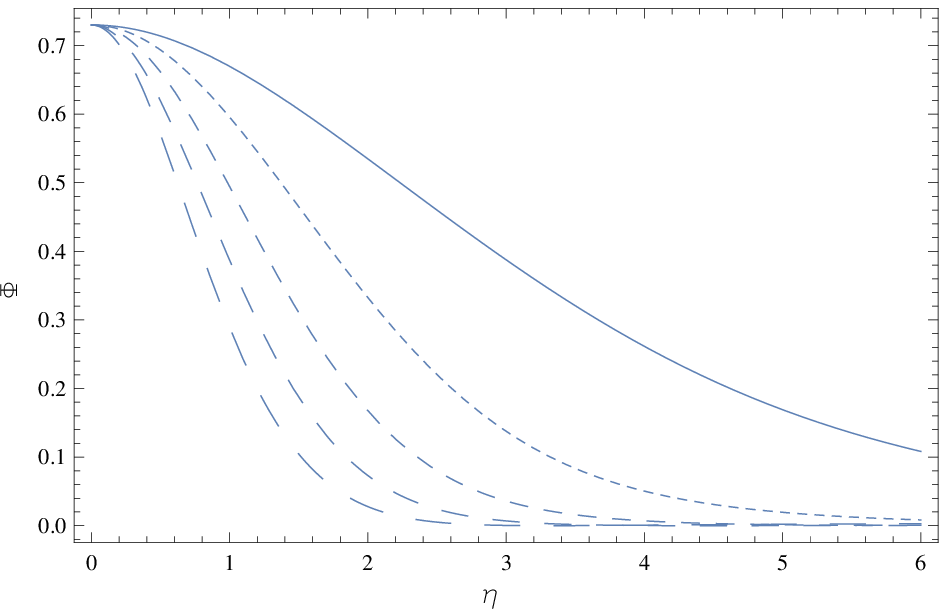}
\hspace{0.5cm}
\includegraphics[width=8.5cm]{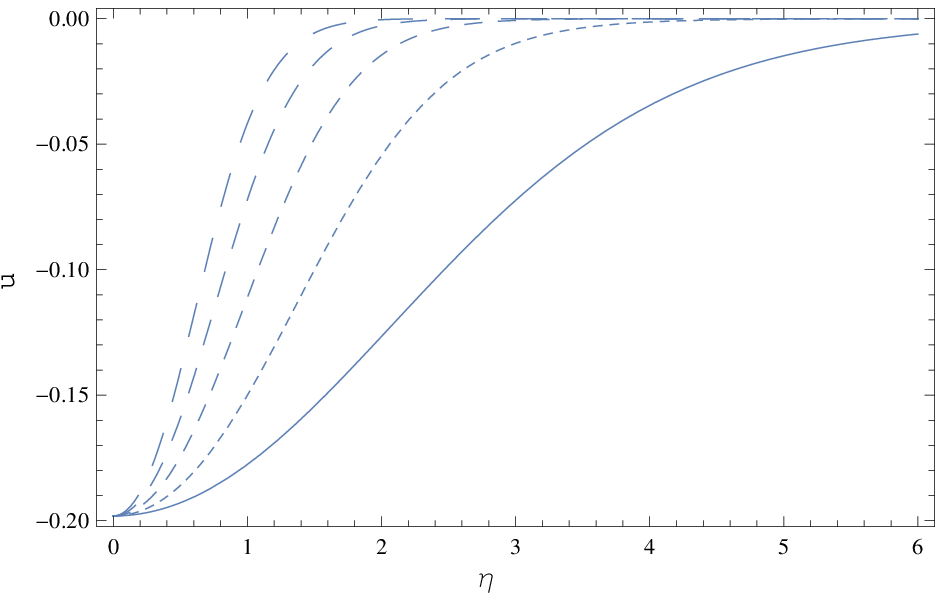}
\caption{Variation of the dimensionless scalar field $\Phi $ (left figure) and of
the Higgs type potential $u$ (right figure) for a stiff fluid star in the
hybrid metric-Palatini gravity theory for $\Phi (0)=0.73$, and for different values of the potential parameters $\mu _0$ and $\xi_0$ and of the central values of the derivatives of the scalar field: $\mu _0=1.05$, $\xi _0=1.15$, $\left.\left(d\Phi /d\eta \right)\right|_{\eta =0}=-1.32\times 10^{-4}$  (solid curve), $\mu _0=2.05$, $\xi _0=1.25$, $\left.\left(d\Phi /d\eta \right)\right|_{\eta =0}=-3 \times 10^{-4}$  (dotted curve), $\mu _0=3.05$, $\xi _0 =1.35$, $\left.\left(d\Phi /d\eta \right)\right|_{\eta =0}=-5.79\times  10^{-4}$
(short dashed curve), $\mu _0=4.05$, $\xi _0=1.45$, $\left.\left(d\Phi /d\eta \right)\right|_{\eta =0}=-9.68\times  10^{-4}$ (dashed curve), and $\mu _0 =5.05$, $\xi _0=1.55$, $\left.\left(d\Phi /d\eta \right)\right|_{\eta =0}=-1.47\times  10^{-3}$ (long dashed
curve), respectively.   The initial conditions for the central density and dimensionless mass used to numerically integrate the hybrid metric-Palatini gravity structure equations are $\theta (0)=1$ and $M_{eff}(0)=0$, respectively.}
\label{f2_stiff}
\end{figure*}

The variations of the scalar field $\Phi $ and of the dimensionless Higgs type potential of the scalar field are represented in Fig.~\ref{f2_stiff}.

 To obtain the variation of the dimensionless potential we have numerically integrated the dimensionless star structure equations (\ref{dmeq})-(\ref{dKGeq}) for the stiff fluid equation of state, with the use of the initial conditions $\theta (0)=1$, $M_{eff}(0)=0$, $\Phi (0)=0.73$, and for different values of the potential parameters $\mu_0$ and $\xi_0$, which generate a set of different central values of the derivative of the scalar field, $\left.\left(d\Phi /d\eta \right)\right|_{\eta =0}=-1.32 \times 10^{-4}$, $\left.\left(d\Phi /d\eta \right)\right|_{\eta =0}=-3 \times 10^{-4}$, $\left.\left(d\Phi /d\eta \right)\right|_{\eta =0}=-5.79 \times 10^{-4}$, $\left.\left(d\Phi /d\eta \right)\right|_{\eta =0}=-9.68\times  10^{-4}$, $\left.\left(d\Phi /d\eta \right)\right|_{\eta =0}=-1.47 \times 10^{-3}$, respectively.   The scalar field $\Phi $, shown in Figs.~\ref{f2_stiff}, is a monotonically decreasing function, reaching the zero value at the surface, a property which is independent on the chosen numerical values of $\mu $ and $\xi $. As one can see from Figs.~\ref{f2_stiff}, for the adopted numerical values of the parameters $\mu _0$ and $\xi _0$, the Higgs type potential of the scalar field takes negative values inside the star, and it reaches the zero value on the star surface.
In order to obtain some definite numerical results we have stopped the numerical integration when the energy density (pressure) at the star's surface reached the value $P\left(\eta _S\right)=0.003$.

As one can see, stiff fluid hybrid metric-Palatini stars are much more massive than their general relativistic counterparts. However, an increase in the central density to $\rho _c=10^{16}$ g/cm$^3$ will decrease the mass of the star to $M=4.24M_{\odot}$, while a central density of $\rho _c=10^{17}$ g/cm$^3$ gives a mass of the hybrid metric-Palatini gravity star of the order of $M=1.34M_{\odot}$ only.

\subsection{Radiation fluid stars}

The radiation fluid is described by the equation of state $P=\theta /3$. The
possibility that stars obeying the radiation equation of state, and
therefore made of photons, could exist, has already been investigated in the literature. Numerical solutions of Einstein's field equation describing
static, spherically symmetric stars made of a photon gas, were obtained in
\cite{Sch}. On the other hand, it was pointed out in \cite{Glen} that a class
of stellar objects called ``Radiation Pressure Supported Stars'' (RPSS) can
exist even in the framework of classical Newtonian gravity. Their generalizations to standard general relativity are denoted ``Relativistic Radiation Pressure Supported Stars'' (RRPSS). It was suggested in \cite{Glen1} that the formation of RRPSSs could take place during the gravitational collapse of massive matter clouds, which may end in a very high density phase. Independently of the details of the contraction process, the trapped radiation flux always reaches the Eddington luminosity at sufficiently large cosmological redshifts $z \gg 1$.

We have obtained the properties of the radiation fluid stars in hybrid metric-Palatini gravity by numerically integrating the star structure equations Eqs.~(\ref{meq}), (\ref{peq}) and (\ref{31}) and (\ref{dmeq})-(\ref{dKGeq}), respectively, for the equation of state $P=\rho c^2/3$ (dimensionless form $P=\theta /3$), respectively. We have used the same values for the parameters of the Higgs potential as in the case of the stiff fluid star. In order to compare the structure of the stars in hybrid metric-Palatini gravity and standard GR we have also presented the corresponding solution of the general relativistic mass continuity and TOV equations (\ref{mgr}) and (\ref{TOVgr}), respectively.

\begin{figure*}[h]
\centering
\includegraphics[width=9.5cm]{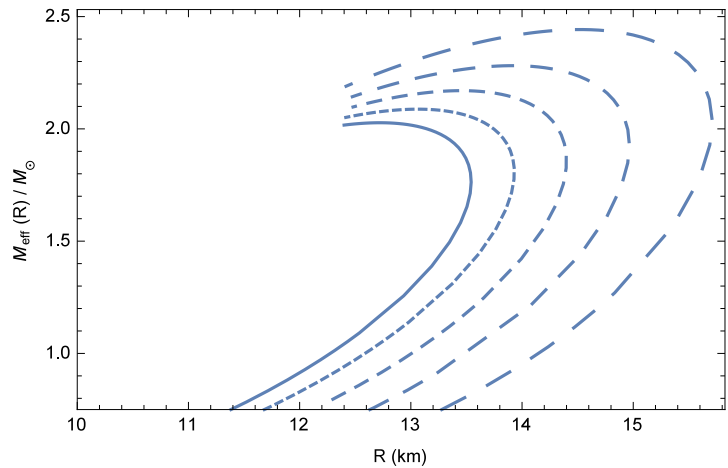}
\caption{Mass-radius relation for radiation fluid stars in hybrid metric-Palatini gravity theory, for $\mu = 10^{-5}\;{\rm cm ^{-1}}$, $\xi = 8.5\times 10^{-10}\;{\rm cm ^{-2}}$ and for different values of $\Phi (0)$ and $\Phi '(0)$: $\Phi \equiv 0$ $\Phi '\equiv 0 $ (standard general relativistic limit) (solid curve), $\Phi (0) = 0.10$, $\Phi '(0) = -2.71 \times 10^{-17}\;{\rm cm}^{-1}$ (dotted curve),  $\Phi (0)=0.15$, $\Phi ' (0)=-1.01 \times 10^{-16}\;{\rm cm}^{-1}$ (short dashed curve), $\Phi (0)=0.20$, $\Phi '(0)=-2.0 \times 10^{-16}\;{\rm cm}^{-1}$ (dashed curve), and $\Phi (0)= 0.25$, $\Phi ' (0)=-2.8 \times 10^{-16}\;{\rm cm}^{-1}$ (long dashed curve).}
\label{f1_rad}
\end{figure*}

To obtain the mass-radius relation we have used the same initial values for the scalar field and its derivative as in the stiff fluid case, and we have varied the central value $\Phi (0)$ of the scalar field. The derivative of the scalar field $\Phi '$ at the center of the star was calculated to be  between $\Phi ' (0) = -2.7\times 10^{-17} \;{\rm cm^{-1}}$ and $\Phi ' (0) = -1.1\times 10^{-16} \;{\rm cm^{-1}}$ for the $\Phi (0)$ values considered. We have stopped the  integration when the density reaches the surface value  $\rho =2\times 10^{14}$ g/cm$^3$. The mass - radius relations for standard general relativistic and hybrid metric-Palatini gravity theory stars are presented in Fig.~\ref{f1_rad}. The central density goes between $3.1\times 10^{14}$ g/cm$^3$ and $2.95\times 10^{15}$ g/cm$^3$ for all curves. For the adopted set of initial conditions and scalar field potential parameters the maximum obtained masses are  $M_{max}=2.027M_{\odot}$, $M_{max}=2.088M_{\odot}$, $M_{max}=2.170M_{\odot}$, $M_{max}=2.281M_{\odot}$,and $M=2.442 M_{\odot}$.

\begin{figure*}[h]
\centering
\includegraphics[width=8.5cm]{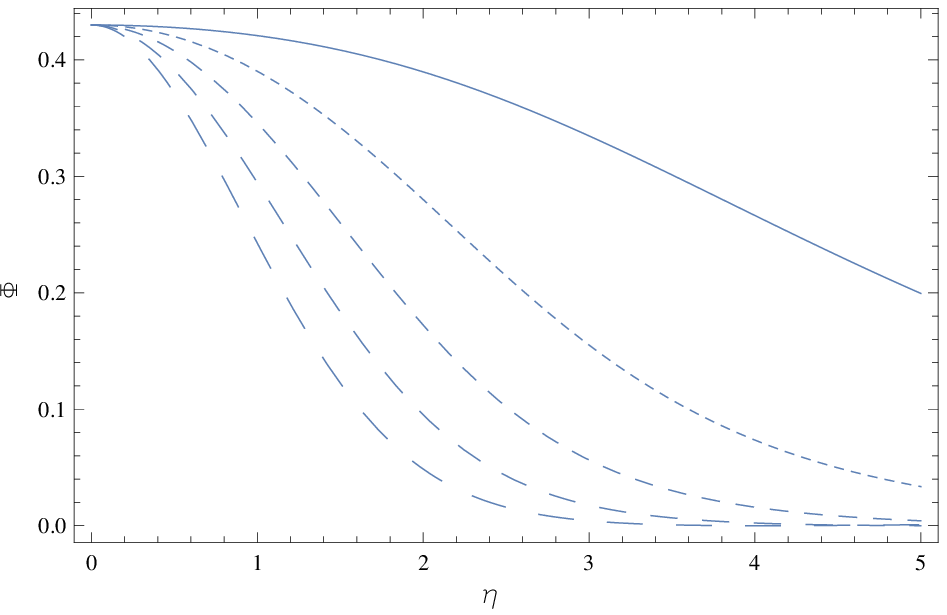}
\hspace{0.5cm}
\includegraphics[width=8.5cm]{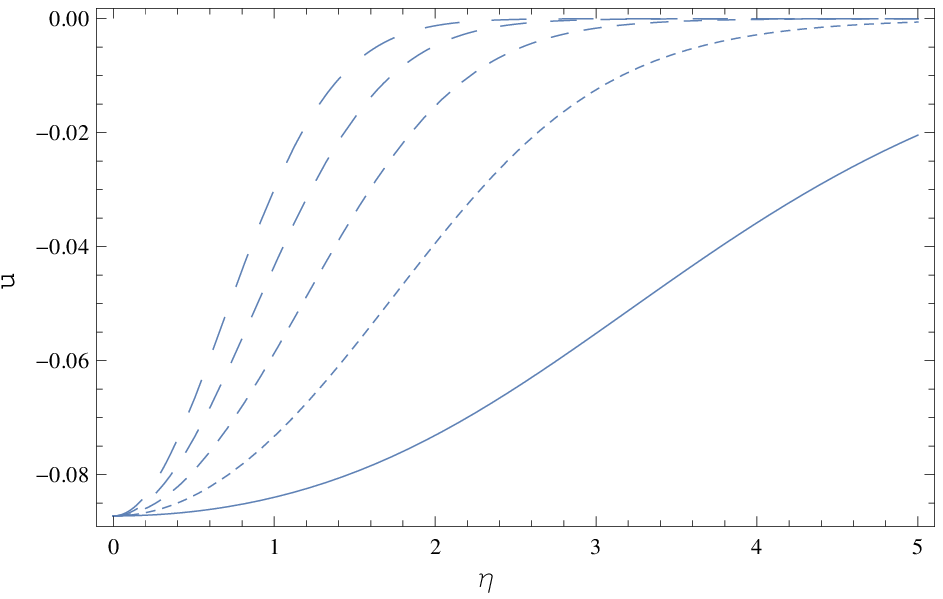}
\caption{Variation of the dimensionless scalar field $\Phi $ (left figure) and of
the Higgs type potential $u$ (right figure) for a radiation fluid star in the
hybrid metric-Palatini gravity theory for different values of the potential parameters $\mu $ and $\xi$: $\mu=1.05$, $\xi =1.15$ (solid curve), $\mu =2.05$, $\xi=1.25$ (dotted curve), $\mu=3.05$, $\xi =1.35$
(short dashed curve), $\mu =4.05$, $\xi =1.45$ (dashed curve), and $\mu =5.05$, $\xi =1.55$ (long dashed
curve), respectively.   The initial conditions used to numerically integrate the hybrid metric-Palatini gravity structure equations are $\theta (0)=1$, $M_{eff}(0)=0$, $\Phi (0)=0.43$, while central values of the derivative of the scalar field, corresponding to the different values of the potential parameters $\mu_0$ and $\xi_0$ are: $\left.\left(d\Phi /d\eta \right)\right|_{\eta =0}=-1.73 \times 10^{-5}$ (solid curve), $\left.\left(d\Phi /d\eta \right)\right|_{\eta =0}=-7.93 \times 10^{-5}$ (dotted curve), $\left.\left(d\Phi /d\eta \right)\right|_{\eta =0}=-1.81\times  10^{-4}$ (short dashed curve), $\left.\left(d\Phi /d\eta \right)\right|_{\eta =0}=-3.24\times  10^{-4}$ (dashed curve), and $\left.\left(d\Phi /d\eta \right)\right|_{\eta =0}=-5.07 \times 10^{-4}$ (long dashed curve), respectively. }
\label{f2_rad}
\end{figure*}

The variations of the scalar field $\Phi $ and of the Higgs type potential of the scalar field for the radiation fluid star are represented in Fig.~\ref{f2_rad}. The curves have been obtained by numerically integrating the set of the dimensionless equations (\ref{dmeq})-(\ref{dKGeq}) by using the initial conditions $\theta (0)=1$, $M_{eff}(0)=0$, $\Phi (0)=0.43$, and where $\left.\left(d\Phi /d\eta \right)\right|_{\eta =0}$ is a function depending on $\Phi (0)$, the potential parameters, and the numerical values of the dimensionless central pressure.

The global structure of the radiation fluid stars in both standard GR and hybrid metric-Palatini gravity is similar to the stiff fluid stars, respectively. The scalar field, presented in Fig.~\ref{f2_rad} is a decreasing function of $\eta $, reaching the value zero on the star surface. The scalar field potential has negative values inside the star, and it vanishes on the surface.
The radiation fluid stars are less massive in both standard GR and hybrid metric-Palatini gravity as compared to the stiff fluid stars. Still, radiation fluid hybrid metric-Palatini stars are much more massive than their general relativistic counterparts.

\subsection{Quark stars}

There are a large number of theoretical arguments suggesting that the
strange quark matter, consisting of $u$, $d$ and $s$ quarks is the most
energetically favorable state of baryon matter \cite{Wi}. There are two ways
for the formation of the strange matter: either the quark-hadron phase
transition in the early Universe, or, alternatively, the conversion of
neutron matter into strange matter inside neutron stars at ultrahigh
densities. The possibility of the existence of stars made of quark matter
was proposed in \cite{It,Bod}. From a theoretical
point of view the equation of state of the quark matter can be derived from
the fundamental Lagrangian of the Quantum Chromodynamics (QCD) \cite{We}. An
important prediction of QCD is the weakening of the quark-quark interaction
at short distances, due to the asymptotic freedom of the theory.

The energy density $\rho $ and the pressure $p$ of a quark-gluon plasma at
temperature $T$ and chemical potential $\mu _{f}$ can be calculated, by
assuming that the interactions of quarks and gluons are sufficiently small,
by using thermal theory. In first order perturbation theory, after
neglecting quark masses, the equation of state of quark matter is given by $%
\rho =\sum_{i=u,d,s,c;e^{-},\mu ^{-}}\rho
_{i}+B$, $p+B=\sum_{i=u,d,s,c;e^{-},\mu ^{-}}p_{i}$, \cite{Wi,Ch},
where $B$, the bag constant, is defined as the difference between the energy
density of the perturbative and non-perturbative Quantum Chromodynamical
vacuum. Therefore the equation of state for quark matter is given by the
Massachusetts Institute of Technology (MIT) bag model equation of state \cite%
{Wi,Ch}
\begin{equation}
p=\frac{1}{3}\left( \rho c^2-4Bc^2\right) .  \label{stateq}
\end{equation}

The equation of state (\ref{stateq}) represents the equation of state of a
gas of massless particles with corrections due to the trace anomaly of
Quantum Chromodynamics, and due to the inclusion of perturbative
interactions. These corrections are always negative, and they reduce the
energy density of the quark-gluon plasma at a given temperature by about a
factor two when the strong interaction coupling constant is of the order of $%
\alpha _{s}=0.5$ \cite{We}. In the dimensionless variables introduced in Eq.~(\ref{dimvar}) the MIT bag model equation of state becomes
\be
P=\frac{1}{3}\left(\theta -4b\right),
\ee
where $b=B/\rho _c$.

\begin{figure*}[h]
\centering
\includegraphics[width=9.5cm]{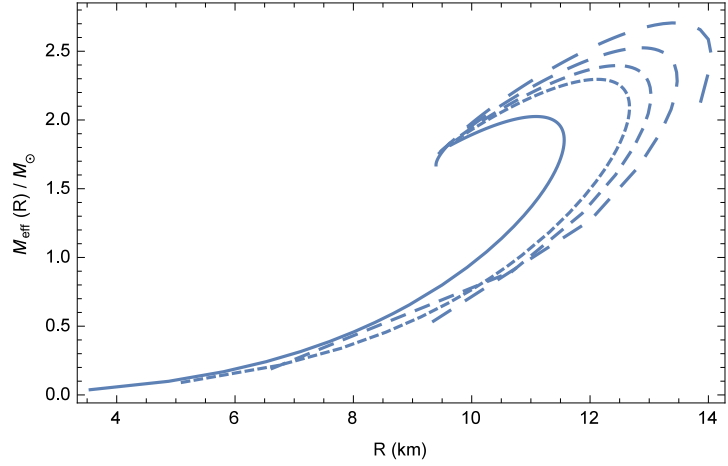}
\caption{Mass-radius relation for quark stars in hybrid metric-Palatini gravity theory, for $\mu = 10^{-5}\;{\rm cm ^{-1}}$, $\xi = 8.5\times 10^{-10}\;{\rm cm ^{-2}}$,  and for different values of $\Phi (0)$ and $\Phi ' (0)$: $\Phi \equiv 0$, $\Phi ' \equiv 0$ (standard general relativistic limit) (solid curve), $\Phi (0) = 0.13$, $\Phi ' (0)=-1.4 \times 10^{-16}\;{\rm cm}^{-1}$ (dotted curve),  $\Phi (0)=0.16$, $\Phi ' (0)=-2.0 \times 10^{-16}\;{\rm cm}^{-1}$ (short dashed curve), $\Phi (0)=0.19$, $\Phi ' (0)=-2.5 \times 10^{-16}\;{\rm cm}^{-1}$ (dashed curve), and $\Phi (0)= 0.22$, $\Phi ' (0)=-2.9 \times 10^{-16}\;{\rm cm}^{-1}$ (long dashed curve).}
\label{f1_quark}
\end{figure*}

In order to obtain the properties of the quark stars in the hybrid metric-Palatini gravity theory, we numerically integrate the star's structure equations (\ref{meq}), (\ref{peq}) and (\ref{31}), and  (\ref{dmeq})-(\ref{dKGeq}), respectively,  by using the MIT bag model equation of state (\ref{stateq}). We have adopted the same values for the parameters of the Higgs potential as in the case of the stiff and radiation fluid stars, respectively. In order to compare the structure of the quark stars in hybrid metric-Palatini gravity and standard GR we have also presented the corresponding solution of the general relativistic mass continuity and TOV equations (\ref{mgr}) and (\ref{TOVgr}). In all cases  integration stops at $P=0$. The mass-radius relations for the quark stars in both standard general relativity and hybrid metric-Palatini gravity are shown in Fig.~\ref{f1_quark}. For the bag constant we have adopted the value $B=10^{14}$ g/cm$^3$. The central density goes between $4.1\times 10^{14}$ g/cm$^3$ and $3.2\times 10^{16}$ g/cm$^3$, respectively. The maximum masses of the quark stars for the chosen values of the parameters of the potential and initial conditions for the scalar field are $M_{max}=2.025M_{\odot}$, $M_{max}=2.296M_{\odot}$, $M_{max}=2.396M_{\odot}$, $M_{max}=2.545M_{\odot}$, and $M_{max}=2.706M_{\odot}$, respectively.

\begin{figure*}[h]
\centering
\includegraphics[width=8.5cm]{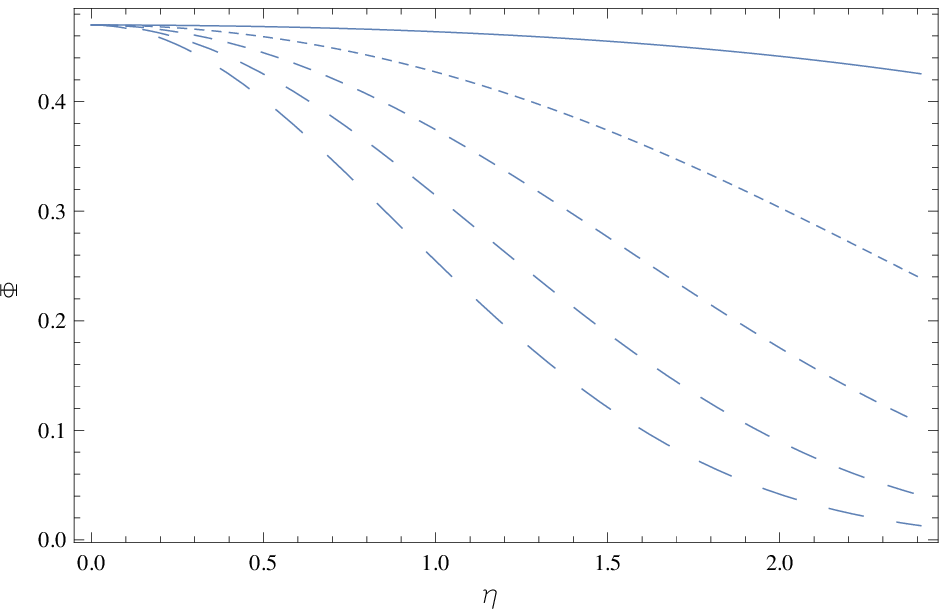}
\hspace{0.5cm}
\includegraphics[width=8.5cm]{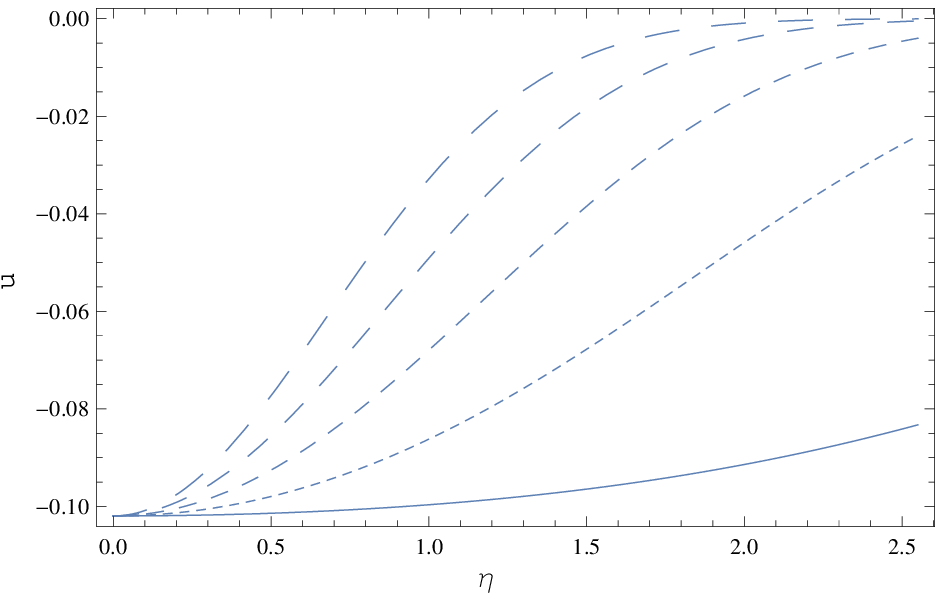}
\caption{Variation of the dimensionless scalar field $\Phi $ (left figure) and of
the Higgs type potential $u$ (right figure) for a quark star in the
hybrid metric-Palatini gravity theory for different values of the potential parameters $\mu $ and $\xi$: $\mu=1.05$, $\xi =1.15$ (solid curve), $\mu =2.05$, $\xi=1.25$ (dotted curve), $\mu=3.05$, $\xi =1.35$
(short dashed curve), $\mu =4.05$, $\xi =1.45$ (dashed curve), and $\mu =5.05$, $\xi =1.55$ (long dashed
curve), respectively.   The initial conditions used to numerically integrate the hybrid metric-Palatini gravity structure equations are $\theta (0)=1$, $M_{eff}(0)=0$, $\Phi (0)=0.43$, while the central values of the derivatives of the scalar field, corresponding to different values of the potential parameters $\mu_0$ and $\xi_0$ are: $\left.\left(d\Phi /d\eta \right)\right|_{\eta =0}=-1.26 \times 10^{-5}$ (solid curve), $\left.\left(d\Phi /d\eta \right)\right|_{\eta =0}=-7.47\times  10^{-5}$ (dotted curve), $\left.\left(d\Phi /d\eta \right)\right|_{\eta =0}=-1.77\times  10^{-4}$ (short dashed curve), $\left.\left(d\Phi /d\eta \right)\right|_{\eta =0}=-3.19\times  10^{-4}$ (dashed curve), and $\left.\left(d\Phi /d\eta \right)\right|_{\eta =0}=-5.026\times  10^{-4}$ (long dashed curve), respectively. For the  bag constant $b$ we have adopted the value $b=0.047$.}
\label{f2_quark}
\end{figure*}

To integrate the dimensionless field equations we have adopted  the initial conditions $\theta (0)=1$, $M_{eff}(0)=0$ and $\Phi (0)=0.47$, and where $\left.\left(d\Phi /d\eta \right)\right|_{\eta =0}$ is a function depending on $\Phi (\eta)$. We have fixed the value of the bag constant as $b=0.047$. The variations with respect to the dimensionless radial coordinate $\eta $ of the scalar field and of the Higgs type potential of the scalar field inside the quark star are presented in Fig.~\ref{f2_quark}.

 The scalar field inside the star, presented in the left plot of Fig.~\ref{f2_quark} is a decreasing function of $\eta $. However, in the case of the quark stars the scalar field does not vanish on the star's surface for most of the adopted values of the parameters of the Higgs potential. On the other hand, similarly to the previous cases, the scalar field potential has negative values inside the star, but generally it does not vanish on the surface.
For the adopted range of the physical parameters, the quark stars are less massive in both standard GR and hybrid metric-Palatini gravity, as compared to the stiff and radiation fluid stars. However, quark hybrid metric-Palatini stars are again much more massive than their general relativistic counterparts. On the other hand the global structure of the quark stars in both standard GR and hybrid metric-Palatini gravity show significant  differences as compared to the stiff and radiation fluid stars, respectively.

\subsection{Bose-Einstein Condensate stars}

Due to the superfluid properties of the neutron matter some compact
astrophysical objects may have a significant part of their matter content in
the form of a Bose-Einstein Condensate \cite{Glen1,Chav}. The
non-relativistic and Newtonian Bose-Einstein gravitational condensate can be
described as a gas, whose density and pressure are related by a barotropic
equation of state $p=p(\rho)$. Generally, the equation of state of the
condensate depends on two physical parameters, the mass of the condensate
particle $m_c$ and the scattering length $a$ \cite{jcap}. In the case of a
condensate with quartic non-linearity, the equation of state is polytropic
with index $n=1$ \cite{Chav,jcap},
\begin{equation}  \label{eqstate}
p\left( \rho \right) =K\rho ^{2},
\end{equation}
with
\begin{equation}
K=\frac{2\pi \hbar ^{2}a}{m_c^{3}} =0.1856\times 10^5 \left(\frac{a}{1\;%
\mathrm{fm}}\right)\left(\frac{m_c}{2m_n}\right)^{-3},
\end{equation}
where $m_n=1.6749\times 10^{-24}$ g is the mass of the neutron. Compact high
density stellar objects having superfluid cores with particles forming
Cooper pairs having masses of the order of two neutron masses, and
scattering length of the order of 10-20 fm, respectively, can have maximum
masses of the order of 2$M_{\odot}$, maximum central density of the order
of $0.1-0.3 \times 10^{16}$ g/cm$^3$, and minimum radii in the range of
10-20 km \cite{Chav}. In the dimensionless variables introduced in Eq.~(\ref{dimvar}) the equation of state (\ref{eqstate}) takes the dimensionless form
\be
P(\theta )=k\theta ^2,
\ee
where $k=K\rho _c/c^2$.

\begin{figure*}[h]
\includegraphics[width=9.5cm]{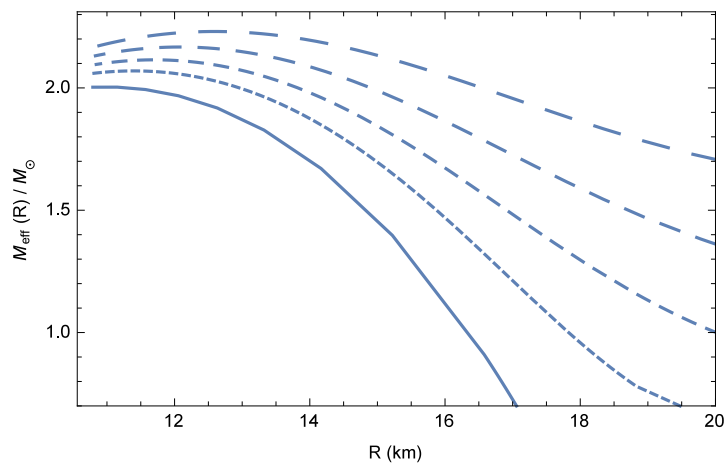}
\caption{Mass-radius relation for Bose-Einstein Condensate stars in hybrid metric-Palatini gravity theory, for $\mu = 10^{-5}\;{\rm cm ^{-1}}$, $\xi = 8.5\times 10^{-10}\;{\rm cm ^{-2}}$, $\Phi ' (0) = -1.8\times 10^{-16}$ to $-5.7 \times 10^{-16} \;{\rm cm^{-1}}$ for the range of the considered central densities, and for different values of $\Phi (0)$: $\Phi \equiv 0$ (standard general relativistic limit) (solid curve), $\Phi (0) = 0.05$ (dotted curve),  $\Phi (0)=0.08$ (short dashed curve), $\Phi (0)=0.11$ (dashed curve), and $\Phi (0)= 0.14$ (long dashed curve).}
\label{f1_Bose}
\end{figure*}

The global properties of the Bose-Einstein Condensate stars in the hybrid metric-Palatini gravity theory have been obtained by numerically integrating the star's structure equations (\ref{meq}), (\ref{peq}) and (\ref{31}), and (\ref{dmeq})-(\ref{dKGeq}), respectively, for the index $n=1$ polytropic equation of state. We have adopted the same values for the parameters $\mu $ and $\xi $ of the Higgs potential as in the case of the stiff, radiation fluid and quark stars, respectively. For the mass of the condensate particle we have adopted the value $m_c=m_n$. In each case the numerical integration stops at $\rho=\rho_c/60$. The central density varies in the range  $2.1\times 10^{13}$ g/cm$^3$ and $6.43\times 10^{15}$ g/cm$^3$ for all cases. In order to compare the global structure of the Boae-Einstein Condensate stars in both hybrid metric-Palatini gravity and general relativity  we have also obtained, and presented, the corresponding numerical solution of the standard general relativistic structure  equations (\ref{mgr}) and (\ref{TOVgr}).  The comparative mass-radius relations for Bose-Einstein Condensate stars in general relativity and hybrid metric-Palatini gravity are presented in Fig.~\ref{f1_Bose}.

The maximum masses obtained for the considered range of parameters are $M_{max}=2.003M_{\odot}$,  $M_{max}=2.070M_{\odot}$, $M_{max}=2.115M_{\odot}$, $M_{max}=2.167M_{\odot}$, and $M_{max}=2.231M_{\odot}$, respectively.

\begin{figure*}[h]
\centering
\includegraphics[width=8.5cm]{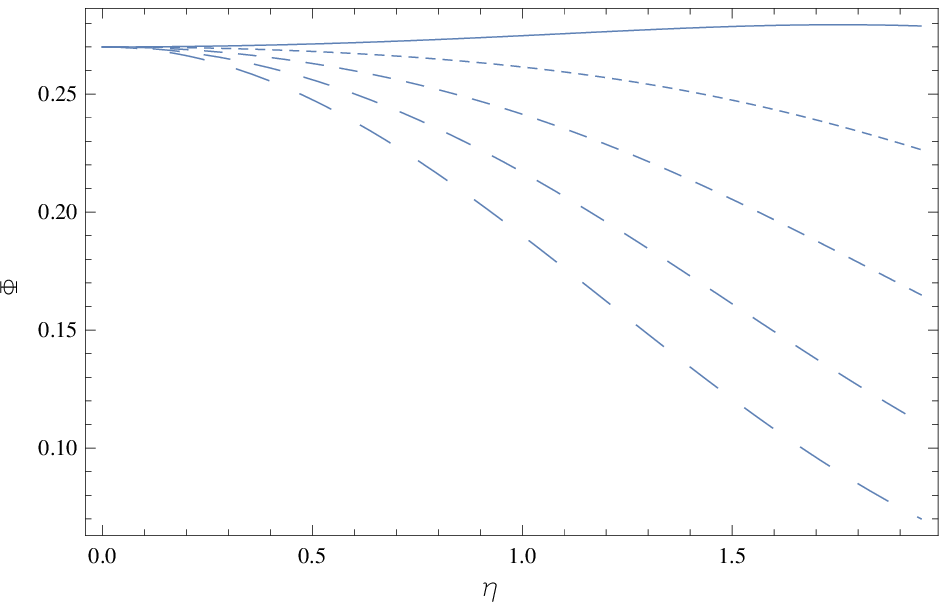}
\hspace{0.6cm}
\includegraphics[width=8.5cm]{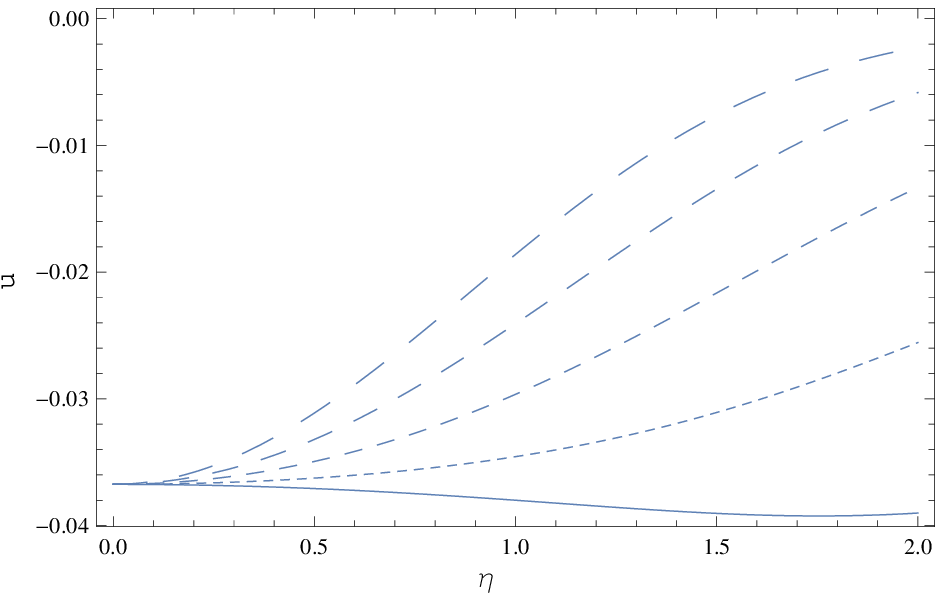}
\caption{Variation of the dimensionless scalar field $\Phi $ (left figure) and of
the Higgs type potential $u$ (right figure) for a Bose-Einstein Condensate star in the
hybrid metric-Palatini gravity theory for different values of the potential parameters $\mu $ and $\xi$: $\mu=1.05$, $\xi =1.15$ (solid curve), $\mu =2.05$, $\xi=1.25$ (dotted curve), $\mu=3.05$, $\xi =1.35$
(short dashed curve), $\mu =4.05$, $\xi =1.45$ (dashed curve), and $\mu =5.05$, $\xi =1.55$ (long dashed curve), respectively. The initial conditions used to numerically integrate the hybrid metric-Palatini gravity structure equations are $\theta (0)=1$, $M_{eff}(0)=0$, $\Phi (0)=0.27$, while the central values of the derivative of the potential, corresponding to different values of the potential parameters $\mu_0$ and $\xi_0$ are: $\left.\left(d\Phi /d\eta \right)\right|_{\eta =0}=1.02\times  10^{-5}$ (solid curve), $\left.\left(d\Phi /d\eta \right)\right|_{\eta =0}=-1.46\times  10^{-5}$ (dotted curve), $\left.\left(d\Phi /d\eta \right)\right|_{\eta =0}=-5.55\times  10^{-5}$ (short dashed curve), $\left.\left(d\Phi /d\eta \right)\right|_{\eta =0}=-1.12\times  10^{-4}$ (dashed curve), and $\left.\left(d\Phi /d\eta \right)\right|_{\eta =0}=-1.85\times 10^{-4}$ (long dashed curve), respectively. For the coefficient  $k$ in the polytropic equation of state we have adopted the value $k=0.1$.}
\label{f2_BEC}
\end{figure*}

In order to integrate the dimensionless set of structure equations for the Bose-Einstein Condensate stars we have adopted the initial conditions  $\theta (0)=1$,  $M_{eff}(0)=0$, $\Phi (0)=0.27$, and where $\left.\left(d\Phi /d\eta \right)\right|_{\eta =0}$ is a function depending on $\Phi (0)$, the potential parameters, and the central pressure. We have fixed the value of the coefficient $k$ in the polytropic equation of state as $k=0.10$,  The variations with respect to the dimensionless radial coordinate $\eta $ of the   scalar field $\Phi $ and of the Higgs type potential of the scalar field for Bose-Einstein Condensate stars are depicted in Fig.~\ref{f2_BEC}.

 The matter pressure (or, equivalently, the energy density) vanishes on the star's surface, which gives the condition $\theta \left(\eta _S\right)=0$, for the determination of the dimensionless radius of the star $\eta _S$. The scalar field inside the star, presented in Fig.~\ref{f2_BEC} has a complex behavior, strongly dependent on the parameters of the Higgs potential. For the first set of numerical parameters, $\Phi$ is an increasing function inside the star, while for the next parameter values it is a decreasing function of $\eta $. Similarly to the case of the quark stars, inside the Bose-Einstein Condensate stars the scalar field does not vanish on the star's surface for the adopted values of the parameters of the Higgs potential. The scalar field potential presents also a complex evolution pattern, with negative values inside the star, and an increasing/decreasing behavior determined by the numerical values of the parameters $\mu $ and $\xi $. For the considered range of parameter values  the scalar field potential does  not vanish on the surface.

For the adopted range of the physical parameters of the Higgs type potential the  Bose-Einstein Condensate stars are less massive in both standard GR and hybrid metric-Palatini gravity, as compared to the stiff, radiation fluid and quark stars, respectively. However, similarly to all previous cases, Bose-Einstein Condensate hybrid metric-Palatini stars are more massive than their general relativistic analogues, but the difference is significantly reduced. Hence the global structure of the Bose-Einstein Condensate  stars in both standard GR and hybrid metric-Palatini gravity show again significant  differences with respect to the stiff and radiation fluid stars, as well as with the quark stars.

\section{Stellar models with fixed functional forms of the scalar field}
\label{sect4}

As a first example of a stellar model in hybrid metric-Palatini gravity in which the form of the scalar field is initially given, we
consider the case in which the scalar field $\Phi $ satisfies the
differential equation
\begin{equation}
f\left( \Phi (\eta )\right) =\frac{d^{2}\Phi }{d\eta ^{2}}+\frac{1}{4}\left(
1+\frac{3}{1-e^{\Phi }}\right) \left( \frac{d\Phi }{d\eta }\right) ^{2}=0.
\label{feq}
\end{equation}
Equation (\ref{feq}) has two solutions, given by
\begin{equation}
\Phi =\Phi _{0}=\mathrm{constant},
\end{equation}%
and
\begin{equation}  \label{form2}
\Phi =\ln \left[ 1+\left( \alpha \eta +\beta \right) ^{4}\right] ,
\end{equation}%
respectively, where
\begin{equation}
\alpha =\frac{1}{4}\frac{e^{\Phi _{0}}}{\left( e^{\Phi _{0}}-1\right) ^{3/4}}%
\Phi _{0}^{^{\prime }},
   \qquad
\beta =\left( e^{\Phi _{0}}-1\right) ^{1/4},
\end{equation}%
and $\Phi _{0}=\Phi \left( 0\right) $, $\Phi _{0}^{^{\prime }}=\left(
d\Phi /d\eta \right) |_{\eta =0}$. In the following we will consider these
two functional forms of $\Phi $,  and we will investigate the physical and geometrical properties of the corresponding stellar models.

\subsection{Effective quark star models-the case $\Phi =\mathrm{constant}$}

In the case $\Phi =\mathrm{constant}$ the generalized dimensionless
Klein-Gordon equation Eq. (\ref{dKGeq}) reduces to the following relation
between the matter density and thermodynamic pressure,
\begin{equation}
P=\frac{1}{3}\left( \theta -4B_{\Phi}\right) ,  \label{bag}
\end{equation}%
where
\begin{equation}
-4B_{\Phi}=e^{\Phi _{0}}\left[ u\left( \Phi \right) -\frac{du\left( \Phi \right) }{%
d\Phi }\right] |_{\Phi =\Phi _{0}}.
\end{equation}

Interestingly enough, the equation of state (\ref{bag}) has the same form as
the MIT bag model equation of state describing quark matter, where $B$
represents the bag constant \cite{Wi}. Hence in the present approach $B_{\Phi}$ can be
interpreted as an effective bag constant, induced by the hybrid
metric-Palatini gravitational theory. In this case the mass continuity and
the hydrostatic equilibrium equations become
\begin{equation}
\frac{dM_{eff}}{d\eta }=\frac{\eta ^{2}}{2}e^{-\Phi _{0}}\left( \theta
+B_{0}\right) ,  \label{q1}
\end{equation}%
\begin{equation}
\frac{d\theta }{d\eta }=-\frac{2e^{-\Phi _{0}}\left( \theta -B\right) \left(
\theta -\alpha \right) \eta ^{2}+3M_{eff}/\eta }{3\eta \left(
1-2M_{eff}/\eta \right) },  \label{q2}
\end{equation}%
where $B_{0}=e^{\Phi _{0}}u\left( \Phi _{0}\right) /2$ and $\alpha =4B_{\Phi}+B_{0}$%
. In the following we will consider that the scalar field potential is of
Higgs type, and, moreover, we assume that for a stable configuration, the scalar field is in
a potential minimum, so that $\left. dV(\phi )/d\phi \right\vert _{\phi
=\phi _{0}}=0$, giving $\phi _{0}=\pm \mu /\sqrt{\xi }$ and $V\left( \phi
_{0}\right) =-\mu ^{4}/4\xi $.

Hence $e^{\Phi _{0}}=1+\phi _{0}=1\pm \mu /\sqrt{\xi }$, and $\left. U\left(
\Phi \right) \right\vert _{\phi =\phi _{0}}=e^{-\Phi }V\left( e^{\Phi
}-1\right) |_{\phi =\phi _{0}}=V/\left(1+\phi _0\right)=-\left( \mu
^{4}/4\xi \right) \left( 1\pm \mu /\sqrt{\xi }\right) $, and we obtain $%
B_{\Phi}=a^{2}\left( \mu ^{4}/4\xi \right) =c^{2}\mu ^{4}/32\pi G_{0}\rho _{c}\xi $%
, $B_{0}=-a^{2}\left( \mu ^{4}/2\xi \right) =-2B_{\Phi}$, and $\theta _{0}=10B_{\Phi}$.

\begin{figure*}[h]
\includegraphics[width=9.5cm]{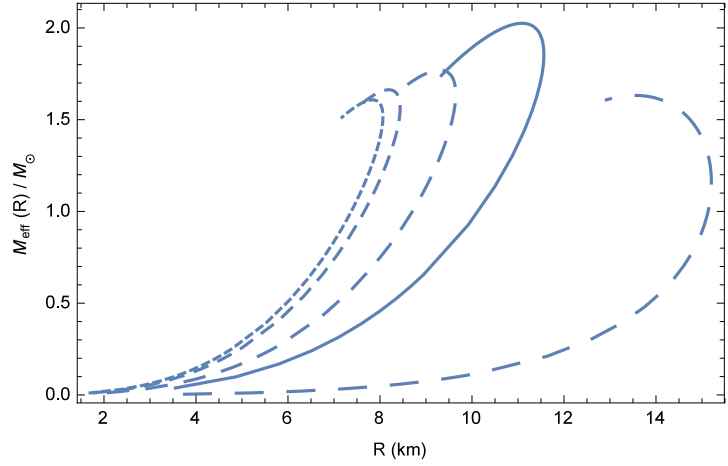}
\caption{Mass-radius relation for hybrid metric-Palatini gravity stars for a constant scalar field, for $\mu = 10^{-5}\;{\rm cm ^{-1}}$, $\xi = 8.5\times 10^{-10}\;{\rm cm ^{-2}}$,  and for different values of $\Phi $: $\Phi \equiv 0$ (standard general relativistic quark star model) (solid curve), $\Phi  = 0.35$ (dotted curve),  $\Phi =0.4$ (short dashed curve), $\Phi =0.45$ (dashed curve), and $\Phi = 0.50$ (long dashed curve).}
\label{f1}
\end{figure*}

The mass-radius relation of the hybrid metric-Palatini star with a constant scalar field are presented, for different values of the effective bag constant $B_{\Phi}$, in Fig.~\ref{f1}. In each case the integration stops at $\rho =4\times 10^{14}$ g/cm$^3$, so that the general relativistic quark star described by the MIT bag model matches the known curve. Central density was varied between $4.1\times 10^{14}$ g/cm$^3$ and $8.5\times 10^{15}$ g/cm$^3$. The maximum masses obtained for the adopted set of parameters are $M_{max}=2.025M_{\odot}$, $M_{\max}=1.608M_{\odot}$, $M_{max}=1.663M_{\odot}$, $M_{max}= 1.770M_{\odot}$, and $M_{max}=1.632M_{\odot}$, respectively.

The effective hybrid metric-Palatini quark star model was compared with the general relativistic quark star model. The maximum mass of the  hybrid metric-Palatini gravity ``analogue'' quark star is much lower than the mass of the ordinary general relativistic quark star, with the constant scalar field in the minimum of the Higgs potential  not giving a significant contribution to the gravitational properties of the system. Of course this conclusion is strongly dependent on the numerical values of the model parameters. By modifying the numerical values of the constant scalar field, ``analogue'' quark star models with different global properties can be constructed.

\subsection{The case $\Phi =\ln \left[ 1+\left( \alpha \eta +\beta \right) ^{4}\right]$}

By adopting for the scalar field $\Phi $ the functional form given by Eq.~(%
\ref{form2}), it follows that the structure equations describing the
interior of a hybrid metric-Palatini gravity star take the form
\begin{equation}\label{fix1}
\frac{dM_{eff}}{d\eta }=-\frac{24\alpha M_{eff}(\alpha \eta +\beta
)^{3}-\eta \left[ 16\alpha (\alpha \eta +\beta )^{3}+2\eta \theta +\eta
u\left( (\alpha \eta +\beta )^{4}+1\right) \right] }{4\left[ (3\alpha \eta
+\beta )(\alpha \eta +\beta )^{3}+1\right] },
\end{equation}%
\begin{equation}\label{fix2}
\frac{d\theta }{d\eta }=-\frac{\theta \left\{ 12M_{eff}\left[ (3\alpha \eta
+\beta )(7\alpha \eta +\beta )(\alpha \eta +\beta )^{2}+1\right] +\eta ^{2}%
\left[ -24\alpha (5\alpha \eta +2\beta )(\alpha \eta +\beta )^{2}+2\eta
\theta -3\eta u\left( (\alpha \eta +\beta )^{4}+1\right) \right] \right\} }{%
3\eta \left( \eta -2M_{eff}\right) \left[ (3\alpha \eta +\beta )(\alpha \eta
+\beta )^{3}+1\right] },
\end{equation}%
\begin{eqnarray}\label{fix3}
\frac{du}{d\eta } &=&\frac{4\left( \eta -2M_{eff}\right) (\alpha \eta +\beta
)}{\eta \left[ (\alpha \eta +\beta )^{4}+1\right] ^{2}}\Bigg\{-12\alpha ^{2}+%
\frac{\eta u(\alpha \eta +\beta )^{2}\left[ \left( \alpha \eta +\beta
\right) ^{4}+1\right] }{\eta -2M_{eff}}+\frac{3\alpha (\alpha \eta +\beta )}{%
\eta \theta \left( \eta -2M_{eff}\right) \left[ (3\alpha \eta +\beta
)(\alpha \eta +\beta )^{3}+1\right] } \times \nonumber  \\
&&\Bigg\{2\eta ^{3}\theta ^{2}+\eta \left[ \eta -2M_{eff}\right] %
\left[ (3\alpha \eta +\beta )(\alpha \eta +\beta )^{3}+1\right] \frac{d\theta}{d\eta }
+\theta \left[ (\alpha \eta +\beta )^{4}+1\right] \left( -8\eta
+12M_{eff}+\eta ^{3}u\right) \Bigg\}\Bigg\} ,
\end{eqnarray}
where to describe the dense matter of the star we have adopted the radiation fluid equation of state $P=\theta /3$. In order to integrate the system of equations (\ref{fix1})-(\ref{fix3}) we need to impose the boundary conditions $M_{eff}(0)=0$, $\theta (0)=1$, $u(0)=u_0$, and $\theta \left(\eta _S\right)=0$, respectively. In the dimensional physical coordinates we represent the scalar field as
\be\label{fixf}
\Phi (r)=\ln\left[1+\left(Ar+C\right)^4\right].
\ee

\begin{figure*}[h]
\includegraphics[width=9.5cm]{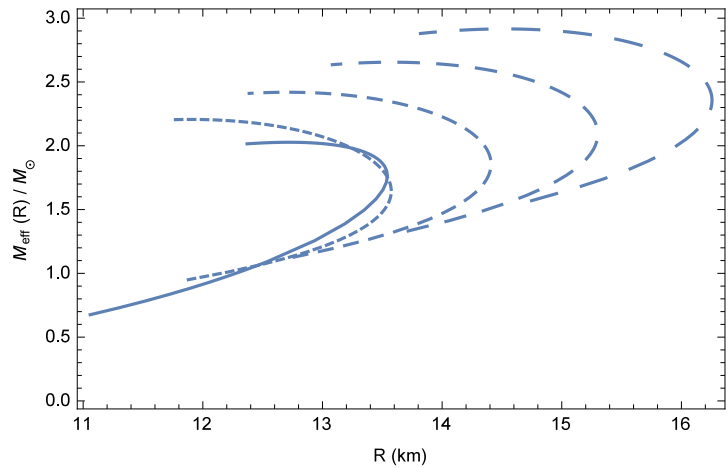}
\caption{Mass-radius relation for hybrid metric-Palatini gravity stars for the scalar field $\Phi (r)=\ln\left[1+\left(Ar+B\right)^4\right]$, for $A=100$ cm $^{-1}$, $U(0)=10^{-3}$ cm$^{-2}$, and for different values of $\Phi (0)$: $\Phi (0) = 2.0$ (dotted curve),  $\Phi (0)=2.1$ (short dashed curve), $\Phi (0)=2.25$ (dashed curve), and $\Phi = 2.3$ (long dashed curve). The solid curve represents the standard general relativistic radiation fluid star model.}
\label{f1_fix}
\end{figure*}

The mass-radius relation for hybrid metric-Palatini gravity stars with the scalar field given by Eq.~(\ref{fixf}) is depicted in Fig.~\ref{f1_fix}. In the numerical computations we have fixed the value of $A$ as $A=100$ cm$^{-1}$, while $C$ was fixed from the initial condition $\Phi (0)=\Phi _0$.  The central density was varied between 0.049 to 0.729. The numerical values of the maximum masses obtained for this set of parameters are $M_{max}=2.027M_{\odot}$ (standard general relativistic model), $M_{max}=2.013M_{\odot}$, $M_{max}=2.206M_{\odot}$, $M_{max}=2.419M_{\odot}$, $M_{max}=2.655M_{|odot}$, and  $M_{max}=2.916M_{\odot}$, respectively.

\begin{figure*}[h]
\centering
\includegraphics[width=8.5cm]{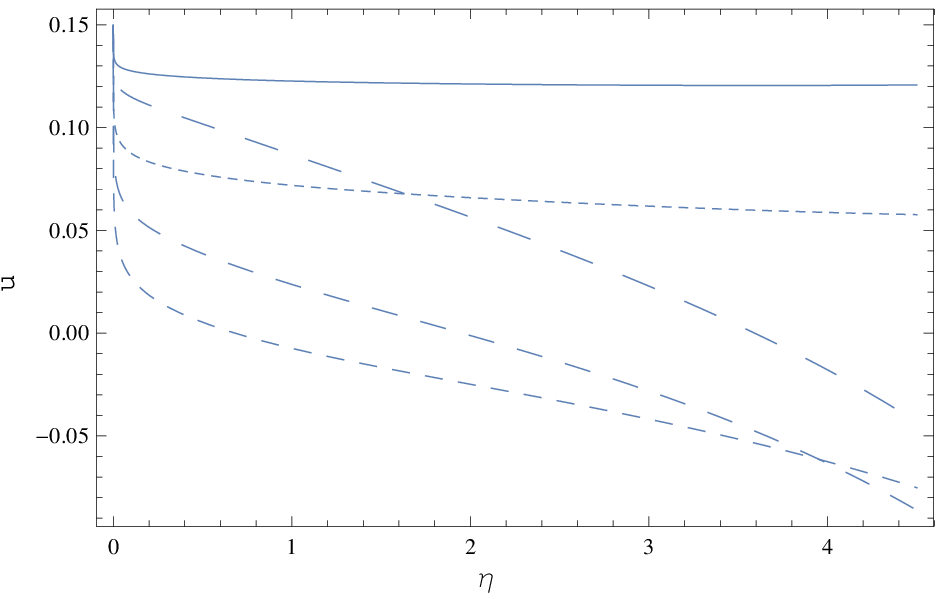}
\hspace{0.5cm}
\includegraphics[width=8.5cm]{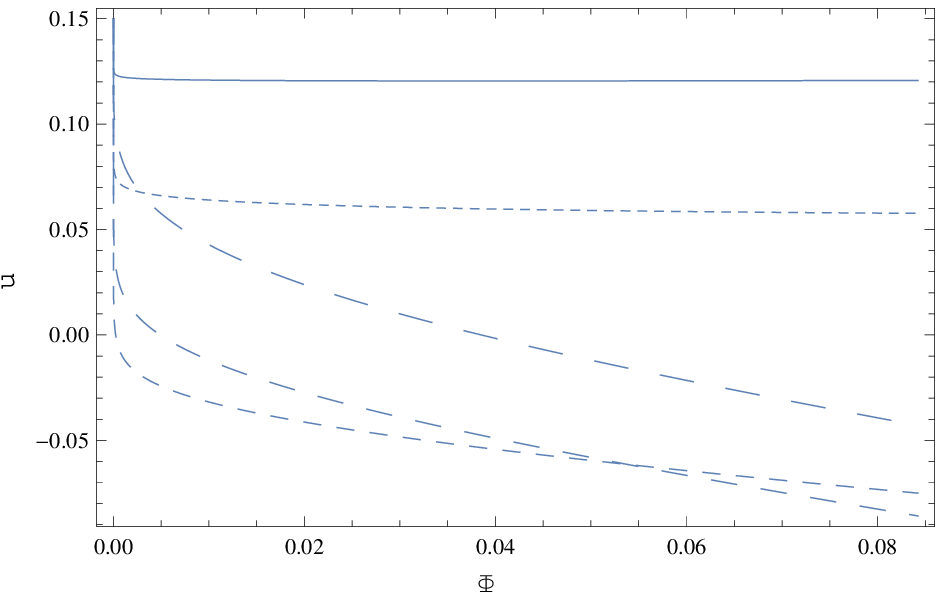}
\caption{Variation of the  scalar field potential $u $ as a function of $\eta $(left figure) and of
the scalar field potential $u$ as a function of $\Phi $ (right figure) for a hybrid metric-Palatini gravity star with $\Phi =\ln \left[ 1+\left( \alpha \eta +\beta \right) ^{4}\right]$, for different values of the parameters $\alpha $ and $\beta $: $\alpha=0.03$, $\beta =0.15$ (solid curve), $\alpha =0.05$, $\beta=0.15$ (dotted curve), $\alpha=0.07$, $\beta =0.15$
(short dashed curve), $\alpha =0.09$, $\beta =0.10$ (dashed curve), and $\alpha =0.11$, $\beta =0.05$ (long dashed
curve), respectively. The dot-dashed curve represents the solution of the structure equations for the radiation fluid star in standard GR.  The initial conditions used to numerically integrate the hybrid metric-Palatini gravity structure equations are $\theta (0)=1$, $M_{eff}(0)=0$,  and $u(0)=0.15$, respectively.}
\label{f2_fix}
\end{figure*}

The variations with respect to $\eta $ of of the scalar field potential $u$, as well as the variation of $u$ with respect to $phi$ are shown in Fig.~\ref{f2_fix}, respectively.


As is transparent from Figs.~\ref{f1_fix}, the hybrid-metric Palatini stars with the scalar field of the form $\Phi =\ln \left[ 1+( \alpha \eta +\beta) ^{4}\right]$ are more massive than their general relativistic counterparts. There is a significant effect of the variation of the scalar field parameters on the effective mass $M_{eff}$ of the star. A small variation in the numerical values of $\alpha $ and $\beta $ determines an important change in the mass of the star. The corresponding masses and radii are also much bigger than those of the general relativistic fluid stars, with masses of the order of two Solar masses. The behavior of the scalar field potential $u$, depicted in Fig.~\ref{f2_fix} is also strongly dependent on the numerical values of $\alpha $ and $\beta $. For small values of $\alpha $ the potential is practically constant inside the star, and it takes only positive values. With the increase of $\alpha $ the potential becomes a monotonically decreasing function of the radial coordinate, also changing sign inside the star. As a function of $\Phi $ the potential shows a similar behavior,  becoming a decreasing function taking negative values outside a small stellar core.

\section{Discussions and final remarks}\label{sect6}
\begin{table}
\begin{center}
 \begin{tabular}{|c |c| c| c|}
 \hline
 Equation of State & $\Phi (0)$ & $M_{max}^{GR}/M_{\odot}$ & $M_{max}^{HMP}/M_{\odot}$ \\
 \hline\hline
 MIT Bag Model & 0.30 & 2.025 & 4.359 \\
 \hline
 Stiff fluid & 0.30 & 3.279 & 3.968 \\
 \hline
 Radiation fluid & 0.30 & 2.256 & 3.660 \\
 \hline
 BEC & 0.30 & 2.230 & 4.971 \\
 \hline
\end{tabular}
\caption{Comparison between the maximum general relativistic $M_{max}^{GR}/M_{\odot}$ and hybrid metric-Palatini $M_{max}^{HMP}/M_{\odot}$ masses obtained for the four equations of state considered in the present study. The parameters of the Higgs type potential used to numerically integrate the hybrid metric-Palatini structure equations are $\mu = 10^{-5}\;{\rm cm ^{-1}}$ and $\xi = 8.5\times 10^{-10}\;{\rm cm ^{-2}}$, respectively, and $\Phi (0) = 0.30 \;{\rm cm^{-1}}$. $\left.\left(d\Phi /dr \right)\right|_{r =0}$ is a function of $\Phi (0)$, of the potential parameters, and of the central densities and pressures. For the MIT bag model and BEC equations of state, the maximum mass occurs at the point of minimum central density.}\label{TableI}
\end{center}
\end{table}

In the present paper, we have investigated the global physical properties of dense compact objects in the hybrid metric-Palatini gravity, which combines elements of the metric and Palatini $f(R)$ theories, and attempts to explain the gravitational phenomena on both local and large scales through a single formalism. An important feature of the theory is the possibility of a scalar-tensor type formulation, which we have used to study the interior of stellar type objects. However, it is important to stress that the gravitational action of the theory differs fundamentally from the Brans-Dicke type action, due to the coupling between the scalar field and the geometry. This coupling generates in static spherically symmetry a rather complicated set of interior field equations, whose solutions can be found only through the intensive use of numerical methods. As a first step in our study, we have derived the basic equations describing the structure of compact objects in hybrid metric-Palatini gravity, namely, the mass continuity equation, the generalized hydrostatic equilibrium equation, and the generalized Klein-Gordon equation, describing the coupling of the scalar field with curvature and matter. An important physical parameter determining the properties of the stars is the self-interaction potential $V$ of the equivalent scalar field. In the present study we have assumed that the potential is of the Higgs type, a choice which is supported by the role such potentials play in elementary particle physics. Other functional forms of the potential (exponential, hyperbolic, power-law etc.) can also be adopted, and they will lead to compact objects having different global properties as compared to those analyzed in this work.

Once the scalar field potential is specified, in order to close the system of structure equations of the star we need to specify either the functional form of the scalar field, or the equation of state of the dense matter. In the framework of the first approach we have investigated two types of solutions of the field equations. In the first case, we have assumed that the scalar field is in the minimum of the Higgs potential, and assumes a constant value. Interestingly enough, this assumption fixes, via the Klein-Gordon equation, the equation of state of the star's matter, which takes the form of the MIT bag model equation of state, which was extensively used to describe the properties of the quark stars. From a simple physical point of view the bag constant forces the quarks to confine into a spherical region of space, with a radius, $r = a$, so that the potential $V(r) = 0$ for $r < a$, with the vacuum pressure $B$ on the bag wall equilibrating the pressure of quarks, and thus stabilizing the hadron. Several mechanisms have been proposed for the formation of quark stars. One  possible scenario is that they may form during the collapse of the core of a massive star after the supernova explosion \cite{Glen1}. Such an explosion may trigger a first or second order phase transition, thus leading to the formation of deconfined quark matter. It has also been pointed out that the core of proto-neutron or neutron stars is a favorable environment for the conversion of neutron matter to quark matter \cite{Ch}. Neutron stars in low-mass X-ray binaries can also accrete enough cosmic matter to undergo a phase transition to become quark stars \cite{Ch}. Hence, the possibility that in hybrid metric-Palatini gravity a phase transition, triggered by the scalar field with Higgs type self-interaction potential, can take place under extreme astrophysical and gravitational conditions (supernova explosions, gamma-ray bursts, accretion etc.) cannot be ruled out. If such a phase transition does occur, the star ends in a minimum of the Higgs potential as a ``true'' or ``analogue'' quark star.

For a given equation of state of the dense matter we have investigated, by numerically integrating the structure equations of the star, four classes of models, corresponding to the stiff fluid, radiation fluid, quark matter and Bose-Einstein Condensate superfluid phase, respectively. In all of these cases we have effectively constructed the hybrid metric-Palatini gravity model of the star, and compared it to its general relativistic counterpart. Our analysis shows that for all these four equations of state the hybrid metric-Palatini gravity stars are much more massive than their standard general relativistic counterparts. For example, for the stiff fluid equation of state, hybrid metric-Palatini stars are about five times heavier than the general relativistic stars. For the same central density quark stars have around two times bigger masses, while superfluid Bose-Einstein Condensate stars are around 1.4 four times more massive. Of course the mass of the star is strongly dependent on its central density, and high central density stars have lower gravitational masses. But the large mass spectrum of the hybrid metric-Palatini stars raises the possibility that stellar mass black holes, with masses in the range of $3.8M_{\odot}$ and $6M_{\odot}$, respectively, could be in fact hybrid metric-Palatini stars (such a possibility was investigated in \cite{Ko} for the case of the quarks stars in the Color-Flavor Locked phase). A comparison of the maximum masses of stellar objects in the hybrid-metric Palatini gravity and of the standard general relativistic values is presented in Table \ref{TableI}.

Many stellar mass black hole candidates have been found recently,  with at least seven of them having masses greater than $5M_{\odot}$. Presently, at least 20 stellar mass black holes have been detected, with masses between 3.8 and 6 Solar masses. However, astronomical estimations give the total number of stellar mass black holes (isolated and in binaries) in our galaxy only to be of the order of 100 millions (see \cite{Ko} and references therein). Therefore the possibility that stellar mass black holes could be ordinary stars dominated by modified gravity effects cannot be ruled out. Hybrid metric-Palatini stars may have higher masses than standard neutron stars, and thus they may be possible stellar mass black hole candidates. A possibility of distinguishing hybrid metric-Palatini stars from standard general relativistic stellar mass black holes could be through the study of the astrophysical properties of the thin accretion disks around rapidly rotating hybrid metric-Palatini stars, and Kerr type black holes, respectively. For such a case we expect that the radiation properties of the accretion disks around general relativistic black holes and modified gravity stars may be different \cite{Koa}. Hence the emission properties of the accretion disk, and of the stars themselves, may be the key signature to differentiate modified gravity stars from ordinary black holes.

 High precision observations of the neutron star mass distribution hae also confirmed the existence of neutron stars with masses of the order of $2M_{\odot}$ \cite{Hor,Demorest, Antoniadis}. One example of such a star is the Black Widow Pulsar B1957+20, an eclipsing binary millisecond pulsar, with the mass estimated to be in the range $1.6-2.4M_{\odot}$ \cite{Kul}. However, a range of 2-2.4 solar masses are very difficult to explain by the standard neutron matter models in the framework of GR, including exotic models like quark or kaon stars. However, these stellar mass values could be easily explained once we model them as hybrid metric-Palatini gravity stars.
Indeed, a hybrid metric-Palatini star exhibits a very complex internal structure, associated with an equally complex stellar dynamics. This is mainly due to the presence of the coupling between the scalar field, geometry and matter. These effects can lead to a number of distinctive astrophysical signatures, which still can make their observational detection to be an extremely difficult task. The possible astrophysical/observational relevance of the hybrid metric-Palatini stars will be considered in a future publication.

\section*{Acknowledgments}

T.H. would like to thank the Institute of Advanced Studies of the Hong Kong University of Science and Technology for the kind hospitality offered during the preparation of this work. B.D. acknowledges financial support from PNIII STAR ACRONIM ASTRES: Centre of Competence For Planetary Sciences, Nr. 118/14.11.2016. F.S.N.L. acknowledges financial support of the Funda\c{c}\~{a}o para a Ci\^{e}ncia e Tecnologia through an Investigador FCT Research contract, with reference IF/00859/2012, and the research grant UID/FIS/04434/2013.



\end{document}